\documentclass[sigconf,manuscript]{acmart}
\renewcommand\footnotetextcopyrightpermission[1]{} 
\usepackage{booktabs}
\usepackage{array}
\usepackage{graphicx}
\usepackage{colortbl}
\usepackage{todonotes}

\AtBeginDocument{%
  \providecommand\BibTeX{{%
    \normalfont B\kern-0.5em{\scshape i\kern-0.25em b}\kern-0.8em\TeX}}}

\copyrightyear{2021}
\acmYear{2021}
\setcopyright{acmlicensed}\acmConference[CHI '21]{CHI Conference on Human Factors in Computing Systems}{May 8--13, 2021}{Yokohama, Japan}
\acmBooktitle{CHI Conference on Human Factors in Computing Systems (CHI '21), May 8--13, 2021, Yokohama, Japan}
\acmPrice{15.00}
\acmDOI{10.1145/3411764.3445188}
\acmISBN{978-1-4503-8096-6/21/05}

\begin{document}

\title{Expanding Explainability: Towards Social Transparency in AI systems}




\author{Upol Ehsan}
\affiliation{%
  \institution{Georgia Institute of Technology}
  \city{Atlanta}
  \state{GA}
  \country{USA}}
 \email{ehsanu@gatech.edu}

\author{Q. Vera Liao}
\affiliation{%
  \institution{IBM Research AI}
  \city{Yorktown Heights}
  \state{NY}
  \country{USA}}
 \email{vera.liao@ibm.com}
 
\author{Michael Muller}
\affiliation{%
  \institution{IBM Research AI}
  \city{Yorktown Heights}
  \state{NY}
  \country{USA}}
 \email{michael_muller@us.ibm.com}
 
 \author{Mark O. Riedl}
\affiliation{%
  \institution{Georgia Institute of Technology}
  \city{Atlanta}
  \state{GA}
  \country{USA}}
 \email{riedl@cc.gatech.edu}
 
\author{Justin D. Weisz}
\affiliation{%
  \institution{IBM Research AI}
  \city{Yorktown Heights}
  \state{NY}
  \country{USA}}
 \email{jweisz@us.ibm.com}

\renewcommand{\shortauthors}{Ehsan, Liao, Muller, Riedl, and Weisz}

\begin{abstract}
 As AI-powered systems increasingly mediate consequential decision-making, their explainability is critical for end-users to take informed and accountable actions. Explanations in human-human interactions are socially-situated. AI systems are often socio-organizationally embedded. However, Explainable AI (XAI) approaches have been predominantly algorithm-centered. We take a developmental step towards socially-situated XAI by introducing and exploring Social Transparency (ST), a sociotechnically informed perspective that incorporates the socio-organizational context into explaining AI-mediated decision-making. To explore ST conceptually, we conducted interviews with 29 AI users and practitioners grounded in a speculative design scenario. We suggested constitutive design elements of ST and developed a conceptual framework to unpack ST’s effect and implications at the technical, decision-making, and organizational level. The framework showcases how ST can potentially calibrate trust in AI, improve decision-making, facilitate organizational collective actions, and cultivate holistic explainability. Our work contributes to the discourse of Human-Centered XAI by expanding the design space of XAI. 
\end{abstract}

\ccsdesc[300]{Human-centered computing~Scenario-based design}
\ccsdesc[500]{Human-centered computing~Empirical studies in HCI}
\ccsdesc[500]{Human-centered computing~HCI theory, concepts and models}
\ccsdesc[300]{Human-centered computing~Collaborative and social computing theory, concepts and paradigms}
\ccsdesc[300]{Computing methodologies~Artificial intelligence}

\keywords{Explainable AI, social transparency, human-AI interaction, explanations, Artificial Intelligence, sociotechnical, socio-organizational context}


\maketitle
\thispagestyle{empty}
\section{Introduction}

Explanations matter. In human-human interactions, they provide necessary delineations of reasoning and justification for one's thoughts and actions, and a primary vehicle to transfer knowledge from one person to another~\cite{lombrozo2012explanation}. Explanations play a central role in sense-making, decision-making, coordination, and many other aspects of our personal and social lives~\cite{hoffman2017explaining}. They are becoming increasingly important in human-AI interactions as well. As AI systems are rapidly being employed in high stakes decision-making scenarios in industries such as healthcare~\cite{loftus2020artificial}, finance~\cite{murawski2019mortgage}, college admissions~\cite{pangburn19schools}, hiring~\cite{dattner2019hiring}, and criminal justice~\cite{hao2019jail}, the need for explainability becomes paramount. Explainability is not only sought by users and other stakeholders to understand and develop appropriate trust of AI systems, but also to support discovery of new knowledge and make informed decisions~\cite{liao2020questioning}. To respond to this emerging need for explainability, there has been commendable progress in the field of Explainable AI (XAI), especially around algorithmic approaches to generate representations of how a machine learning (ML) model operates or makes decisions.

Despite the recent growth spurt in the field of XAI, studies examining how people actually interact with AI explanations have found popular XAI techniques to be ineffective~\cite{alqaraawi2020evaluating,poursabzi2018manipulating,zhang2020effect}, potentially risky~\cite{kaur2020interpreting,stumpf2016explanations}, and underused in real-world contexts~\cite{liao2020questioning}. The field has been critiqued for its techno-centric view, where ``inmates [are running] the asylum''~\cite{miller2019explanation}, based on the impression that XAI researchers often develop explanations based on their own intuition rather than the situated needs of their intended audience. Currently, the dominant algorithm-centered XAI approaches make up for only a small fragment of the landscape of explanations as studied in the Social Sciences~\cite{miller2019explanation,lombrozo2012explanation,wang2019designing,mittelstadt2019explaining} and exhibit significant gaps from how explanations are sought and produced by people. Certain techno-centric pitfalls that are deeply embedded in AI and Computer Science, such as Solutionism (always seeking technical solutions) and Formalism (seeking abstract, mathematical solutions)~\cite{selbst2019fairness, green2020algorithmic}, are likely to further widen these gaps.

One way to address the gaps would be to critically reflect on the status quo. Here, the lenses of Agre's Critical Technical Practice (CTP) ~\cite{agre1997toward,agre1997computation} can help. CTP encourages us to question the core epistemic and methodological assumptions in XAI, critically reflect on them to overcome impasses, and generate new questions and hypotheses. By bringing the unconscious aspects of experience to our conscious awareness, critical reflection makes them actionable~\cite{sengers2005reflective,dourish2004action,dourish2004reflective}. Put differently, a CTP-inspired reflective perspective on XAI~\cite{ehsan2020human} will encourage us to ask: by continuing the dominant algorithm-centered paradigm in XAI, what perspectives are we missing? How might we incorporate the marginalized perspectives to embody alternative technology? In this case, a dominant XAI approach can be construed as algorithm-centered that privileges technical transparency and circumscribes the epistemic space of explainable AI around model transparency. An algorithm-centered approach can be effective if explanations and AI systems existed in a vacuum. However, it is not the case that explanations and AI systems are devoid of situated context.

On one hand, explanations (as a construct) are socially situated~\cite{lombrozo2011instrumental, lombrozo2012explanation, wilkenfeld2015inference, miller2019explanation}. Explanation is first and foremost a shared meaning-making process that occurs between an explainer and an explainee. This process is dynamic to the goals and changing beliefs of both parties~\cite{dennett1989intentional,hume2000enquiry,hilton1996mental,heider1958psychology}. For our purposes in this paper, we adopt the broad definition that an explanation is an answer to a \textit{why}-question~\cite{miller2019explanation,dennett1989intentional,lewis1986causal}.

On the other hand, implicit in AI systems are \textit{human-AI assemblages}. Most consequential AI systems are deeply embedded in socio-organizational tapestries in which groups of humans interact with it, going beyond a 1-1 human-AI interaction paradigm. Given this understanding, we might ask: if both AI systems and explanations are socially-situated, then why are we not requiring incorporation of the social aspects when we conceptualize explainability in AI systems? How can one form a holistic understanding of an AI system and make informed decisions if one only focuses on the technical half of a sociotechnical system?

We illustrate the shortcomings of a solely technical view of explainability in the following scenario, which is inspired by incidents described by informants in our study.

\begin{quote}
	\textit{You work for a leading cloud software company, responsible for determining product pricing in various markets. Your institution built a new AI-powered tool that provides pricing recommendations based on a wide variety of factors. This tool has been extensively evaluated to assist you on pricing decisions. One day, you are tasked with creating a bid to be the cloud provider for a major financial institution. The AI-powered tool gives you a recommended price. You might think, why should I trust the AI’s recommendation? You examine a variety of technical explanations the system provides: visualizations of the model's decision-making process and descriptions of how the algorithm reached this specific recommendation. Confident at the soundness of the model's recommendation, you create the bid and submit it to the client. You are disheartened to learn that the client rejected your bid and instead accepted the bid from a competitor.}
\end{quote}

Given a highly-accurate machine learning model, along with a full complement of technical explanations, why should the seller's pricing decision not have been successful? It is because the answer to the \textit{why}-question is not limited to the machine explaining itself. It is also in the situational and socio-organizational context, which one can learn from how price recommendations were handled by other sellers. What other factors went into those decisions? Were there regulatory or client-specific (e.g., internal budgetary constraints) issues that were beyond the scope of the model? Did something drastic happen in the operating environment (e.g., a global pandemic) that necessitated a different strategy? In other words, situational context matters and it is with this context the ``why'' questions could be answered effectively and completely.

At a first glance, it may seem that socio-organizational context has nothing to do with explaining an AI system. Therein lies the issue --- where we draw the boundary of our epistemic canvas for XAI matters. If the boundary is traced along the bounds of an algorithm, we risk excluding the human and social factors that significantly impact the way people make sense of a system. Sense-making is not just about opening the closed box of AI, but also about who is around the box, and the sociotechnical factors that govern the use of the AI system and the decision. Thus the ``ability'' in explainability does not lie exclusively in the guts of the AI system~\cite{ehsan2020human}. For the XAI field as a whole, if we restrict our epistemic lenses to solely focus on algorithms, we run the risk of perpetuating the aforementioned gaps,  marginalizing the human and sociotechnical factors in XAI design. The lack of incorporation of the socio-organizational context is an epistemic blind spot in XAI. By identifying and critically reflecting on this epistemic blind spot, we can begin to recognize the poverty of algorithm-centered approaches.

In this paper, we address this blind spot and expand the conceptual lens of XAI by reframing explainability beyond algorithmic transparency, focusing our attention to the human and socio-organizational factors around explainability of AI systems. Building upon relevant concepts that promote transparency of social information in human-human interactions, we introduce and explore Social Transparency (ST) in AI systems. Using a scenario-based design, we create a speculative instance of AI-mediated decision-support system and use it to conduct a formative study with 29 AI users and practitioners. Our study explores whether and how proposed constitutive design elements address the epistemic blind spot of XAI -- incorporating socio-organizational contexts into explainability. We also investigate whether and how ST can facilitate AI-mediated decision-making and other user goals. This paper is not a full treatise of how to achieve socially-situated XAI; rather a first step toward that goal by operationalizing the concept in a set of design elements and considering its implications for human-AI interaction. In summary, our contributions are fourfold:

\begin{itemize}
    \item We highlight an epistemic blind spot in XAI -- a lack of incorporation of socio-organizational contexts that impact the explainability of AI-mediated decisions -- by using a CTP-inspired reflective approach to XAI.
    \item We explore the concept of Social Transparency (ST) in AI systems and develop a scenario-based speculative design that embodies ST, including four categories of design features that reflect \textit{What}, \textit{Why}, \textit{Who}, and \textit{When} information of past user interactions with AI systems.
    \item We conduct a formative study and empirically derive a conceptual framework, highlighting three levels of context around AI-mediated decisions that are made visible by ST and their potential effects: technological (AI), decision, and organizational contexts.
    \item We share design insights and potential challenges, risks, and tensions of introducing ST into AI systems.
\end{itemize}

\section{Related work}
We begin with a in-depth review of related work in XAI field, further highlighting the danger of the epistemic blind spot. We then discuss a shift in broader AI related work towards sociotechnical perspectives. Lastly, we review work that pushed towards transparency of socio-organizational contexts in human-human interactions.

\subsection{Explainable AI (XAI)}

Although there is no established consensus on the complete set of factors that makes an AI system explainable, XAI work commonly shares the goal of making an AI system's functioning or decisions \textit{easy to understand} by people~\cite{lipton2018mythos,arrieta2020explainable,gilpin2018explaining,miller2019explanation,gunning2017explainable,ras2018explanation,carvalho2019machine, ehsan2019automated}. Recent work also emphasizes that explainability is an audience-dependant instead of a model-inherent property~\cite{arrieta2020explainable,mohseni2018multidisciplinary,arya2019one,miller2019explanation, ehsan2020human}. Explainability is often viewed more broadly than model transparency or intelligibility~\cite{gilpin2018explaining,ras2018explanation,lipton2018mythos}. For example, a growing research area of XAI focuses on techniques to generate \textit{post-hoc} explanations~\cite{ehsan2019automated}. Instead of directly elucidating how a model works internally, post-hoc explanations typically justify an opaque' model's decision by rationalizing the input and output or providing similar examples. Lipton discussed the importance of post-hoc explanations to provide useful information for decision makers, and its similarity with how humans explain~\cite{lipton2018mythos}. At a high level, Gilpin et al.~\cite{gilpin2018explaining} argued that the transparency of model behaviors alone is not enough to satisfy the goal of ``\textit{gain[ing] user trust or produc[ing] insights about the cause of the decisions},'' but rather, explainability requires other capabilities such as providing responses to user questions and the ability to be audited.

Since an explanation is only explanatory if it can be consumed by the recipient, many recognize the importance of taking user-centered approaches to XAI~\cite{miller2019explanation,shneiderman2020human,vaughan20201}, and the indispensable role that the HCI community should play in advancing the field. While XAI has experienced a recent surge in activities, the HCI community has a long history of developing and studying explainable systems, such as explainable recommender systems, context-aware systems, and intelligent agents, as outlined by Abdul et al.~\cite{abdul2018trends}.  Moreover, XAI's disconnect with the philosophical and psychological grounds of human explanations has been duly noted~\cite{mittelstadt2019explaining}, as best represented by Miller's call for leveraging insights from the Social Sciences~\cite{miller2019explanation}.  Wang et al. reviewed decision-making theories and identified many gaps in XAI output to support the complete cognitive processes of human reasoning~\cite{wang2019designing}. From these lines of work, we highlight a few critical issues that are most relevant to our work.

First, there is a dearth of user studies and a lack of understanding on how people actually perceive and consume AI explanations~\cite{doshi2017towards,vaughan20201}. Only until recently have researchers began to conduct controlled lab studies to rigorously evaluate popular XAI techniques~\cite{buccinca2020proxy,cai2019effects,dodge2019explaining,poursabzi2018manipulating,cheng2019explaining,lai2020chicago, ehsan2019automated}, as well as studies to understand real-world user needs for AI explainability~\cite{liao2020questioning,kaur2020interpreting,hong2020human}.  Accumulating evidence shows that XAI techniques are not as effective as assumed. There have been rather mixed results on whether current XAI techniques could appropriately enhance user trust~\cite{cheng2019explaining,poursabzi2018manipulating,yang2020visual} or the intended task performance, whether for decision making~\cite{liao2020questioning,zhang2020effect,buccinca2020proxy}, model evaluation~\cite{cai2019effects,alqaraawi2020evaluating,dodge2019explaining}, or model development~\cite{kaur2020interpreting}. For example, Alqarrawi et al. evaluated the effectiveness of saliency maps~\cite{alqaraawi2020evaluating} -- a popular explanation technique for image classification models -- and found they provided very limited help for evaluating the model. Kauer et al. studied how data scientists use popular model interpretability tools and found them to be frequently misused~\cite{kaur2020interpreting}. Liao et al. interviewed practitioners designing AI systems and reported their struggle with popular XAI techniques due to a lack of actionability for end users. Recent studies also reported detrimental effects of explanations for AI system users including inducing over-trust or over-estimation of model capabilities~\cite{kaur2020interpreting, stumpf2016explanations,smith2020no}, and increasing cognitive workload~\cite{abdul2020cogam,Ghai2020XAL}. Moreover, while XAI is often claimed to be a critical step towards accountable AI, empirical studies have found little evidence that explanations improve a user's perceived accountability or control over AI systems~\cite{rader2018explanations,smith2020no}.

Second, in human reasoning and learning, explanation is both a \textit{product} and a \textit{process}. In particular, it is a \textit{social process}~\cite{miller2019explanation} as part of
a conversation or social interaction. Current technical XAI work typically takes a product-oriented view by generating a representation of a model's internals~\cite{lombrozo2012explanation}. However, explanations are also sought first and foremost as a knowledge transfer process from an explainer to an explainee. A process-oriented view has at least two implications for XAI. First, the primary goal of explanation should be to enable the explainee to gain knowledge or make sense of a situation or event, which may not be limited to a model's internals. Second, as a transfer of knowledge, explanations should be presented relative to the explainee's beliefs or knowledge gaps~\cite{miller2019explanation}. This emphasis on tailoring explanation according to explainee's knowledge gaps has been a focus of prior HCI work on explainable systems~\cite{lim2009and,lim2010toolkit,lim2019these,liao2020questioning}. Recent work has also begun to explore interactive explanations that could address users' follow-up questions as a way to fill individual knowledge gaps~\cite{weld2019challenge,spinner2019explainer}.  However, sometimes these knowledge gaps lie outside of the system, which may require providing information that is not related to its internal mechanics~\cite{abdul2018trends}.

Finally, we argue that AI systems are socially situated, but sociotechnical perspectives are mostly absent in current XAI work. One recent study by Hong et al.~\cite{hong2020human} investigated how practitioners view and use XAI tools in organizations using ML models. Their findings suggest that the process of interpreting or making sense of an AI system frequently involves cooperation and mental model comparison between people in different roles, aimed at building trust not only between people and the AI system, but also between people within the organization~\cite{hong2020human}. Our work builds on these observations, as well as prior work on sociotechnical approaches to AI systems which we review below.

\subsection{Sociotechnical approaches to AI}

Our work is broadly motivated by work on sociotechnical approaches to AI. Academia and society at large have begun to recognize the detrimental effect of a techno-centric view on AI~\cite{shneiderman2020human,vaughan20201,sabanovic2010robots}. Since AI systems are socially situated, their development should carefully consider social, organizational, and cultural factors that may govern their usage. Otherwise one may risk deploying an AI system un-integrated into individual and organizational workflows~\cite{makarius2020rising,wolf2019evaluating}, potentially resulting in misuse, mistrust~\cite{yang2016investigating,yang2019unremarkable}, or having profound ethical risks and unintended consequences, especially for marginalized groups~\cite{mohamed2020decolonial,suresh2019framework,sanchez2020does}.

Researchers have proposed ways to make AI systems more human-centered and sensitive to socio-organizational contexts. 
Bridging rich veins of work in AI, HCI, and critical theory, such as Critical Technical Practices~\cite{agre1997computation} and Reflective Design~\cite{sengers2005reflective}, Ehsan and Riedl delineate the foundations of a Reflective Human-centered XAI (HCXAI). \textit{Reflective HCXAI} is a sociotechnically informed perspective on XAI that is critically reflective of dominant assumptions and practices of the field~\cite{ehsan2020human}, and sensitive to the values of diverse stakeholders, especially marginalized groups, in its proposal of alternative technology. Zhu et al. proposed Value Sensitive Algorithm Design~\cite{zhu2018value} by engaging stakeholders in the early stages of algorithm creation, to avoid biases in design choices or compromising stakeholder values. Several researchers have leveraged design fictions and speculative scenarios to elicit user values and cultural perspectives for AI system design~\cite{cheon2016integrating,cheon2018futuristic,muller2017exploring}. Šabanovic developed a framework of Mutual-Shaping and Co-production~\cite{sabanovic2010robots} by involving users in the early stages of robot design and engaging in reflexive practices.  Jones et al~\cite{jones2013design} proposed a design process for intelligent sociotechnical systems with equal attention to analysis of social concepts in the deployment context and representing such concepts in computational forms.

More fundamentally, using a Science and Technology Studies (STS)  lens~\cite{suchman1987plans}, scholars have begun critically reflecting on the underlying assumptions made by AI algorithmic solutions. Mohamed et al.~\cite{mohamed2020decolonial} examined the roles of power embedded in AI algorithms, and suggested applying decolonial approaches to enable AI technologies to center on vulnerable groups that may bear negative consequences of technical innovation. Green and Viljoen~\cite{green2020algorithmic} diagnosed the dominant mode of AI algorithmic reasoning as ``algorithmic formalism'' -- an adherence to prescribed form and rules -- which could lead to harmful outcomes such as reproducing existing social conditions and a technologically-deterministic view of social changes. The authors pointed out that addressing these potential harms requires attending to the internal limits of algorithms and the social concerns that fall beyond the bounds of algorithmic formalism. In the context of fair ML, Selbst et al.\cite{selbst2019fairness} questioned the implications of algorithmic abstraction that are essential to ML. Abstracting away the broader social context can cause AI technical interventions to fall into a number of traps: Framing, Portability, Formalism, Ripple Effect, and Solutionism. The authors suggested to mitigate these problems by extending abstraction boundaries to include social factors rather than purely technical ones. In a similar vein, field work on algorithmic fairness often found that meaningful interventions toward usable and ethical algorithmic systems are non-technical, and that user community derive most value from localized, as opposed to ``scalable'' solutions~\cite{katell2020toward,lee2019webuildai}.

Our work is aligned with and builds on these views obtained through the sociotechnical lens. These perspectives inform our thinking as we expand the boundaries of XAI to include socio-organizational factors, and challenge a formalist perspective that peoples' meaning-making processes could be resolved through algorithmic formalisms. Our work takes an operational step towards sociotechnical XAI systems by expanding the design space with ST.

\subsection{Social transparency and related concepts} 

Our work is also informed by prior work that studied social transparency and related concepts in human-human interactions. The concept of making others' activities transparent plays a central role in HCI and Computer-Supported Cooperative Work (CSCW) literature~\cite{suchman1995making,star1999layers}. Erickson and Kellogg proposed the concept of and design principles for Social Translucence, in which ``social cues'' of others' presence and activities are made visible in digital systems, so that people can apply familiar social rules to facilitate effective online communication and collaboration. Gutwin et al.'s seminal work on group awareness~\cite{gutwin2002descriptive} for groupware supporting distributed teams provides an operational design framework. It sets out elements of knowledge that constitute group awareness, including knowledge regarding \textit{Who}, \textit{What}, and \textit{Where} to support awareness related to the present, and \textit{How}, \textit{When}, \textit{Who}, \textit{Where}, and \textit{What} for awareness related to the past. Theses theories have since inspired a bulk of work that created new design features and design spaces for social and collaborative technologies (e.g.~\cite{erickson2003social,mcdonald2012building,gilbert2012designing,gutwin2004group,kim2020enriched}).

Building upon social translucence and awareness, Stuart et al.~\cite{stuart2012social} conceptualized Social Transparency (ST) in networked information exchange. In particular, it extends the visibility of one's direct partner and the effect on their dyadic interactions, to also encompass one's role as an \textit{observer} of others' interactions made visible in the network. Their framework describes three social dimensions made visible to people by ST: identity transparency, content transparency, and interaction transparency. This framework then considers a list of \textit{social inferences} people could make based on these visible dimensions (e.g. perceived similarity and accountability based on identity transparency; activity awareness based on content transparency; norms and social networks based on interaction transparency), and their second order effects for the groups or community. Social transparency theory has been used to design and analyze various social media features and their impact on social learning~\cite{dabbish2012social,nguyen2015perverse}, social facilitation~\cite{huang2013don}, and reputation management~\cite{dabbish2012social}.

The above work focused on how ST -- making others' activities visible -- affects collaboration and cooperative behaviors with other people. Our work also draws on two other important aspects that ST could potentially support for decision-making. One is on knowledge sharing and acquisition. As reviewed by Ackerman et al,~\cite{ackerman2013sharing}, CSCW systems supporting organizational knowledge management fall into two categories: a repository model that externalizes peoples' knowledge as sharable artifacts or objects; and an expertise-sharing model that supports locating the appropriate person to have in-situ access to knowledge. The CSCW community's shift from the former to the latter category represents a shift of emphasis from \textit{explicit} to \textit{tacit} knowledge. Transparency of others' communications could facilitate expertise location through the acquisition of organizational meta-knowledge (e.g., \textit{who knows what} and \textit{who knows whom}), as a type of ``ambient awareness'' coined by Leonardi in the context of enterprise social media~\cite{leonardi2014social,leonardi2015ambient}.
This position is also related to the development of \textit{Transactive Memory Systems (TMS)}~\cite{yoo2001developments,moreland2006transactive,brandon2004transactive,nardi2002networkers} that relies on meta-knowledge to optimize the storage and retrieval of knowledge across different individuals.  A sufficiently fluent TMS can evolve to a form of team cognition of or ``collective mind''~\cite{weick1993collective,hutchins1991social} that can lead to better collective performance~\cite{hollingshead2003potential,austin2003transactive}.

Social transparency could also guide or validate peoples' judgment and decision as cognitive heuristics. Cognitive heuristics are a key concept in decision-making~\cite{kahneman1982judgment}, which refers to ``rules of thumb'' people follow to quickly form judgments or find solutions to complex problems. By making visible what other people selected, interacted with, approved or disapproved, ST could invoke many social and group-based heuristics such as bandwagon or endorsement heuristics (following what many others' do), authority or reputation heuristics (following authority),  similarity heuristics (following people in similar situations), and social presence heuristics (favoring a social entity over a machine)~\cite{sundar2008main,metzger2010social,metzger2013credibility}. How these ST-rendered heuristics affect peoples' decisions and actions has been studied in a wide range of technologies such as reputation systems~\cite{resnick2000reputation} and social media. In particular, they play a critical role in how people evaluate the trustworthiness, credibility, and agency of technologies~\cite{sundar2008main,metzger2010social,metzger2013credibility}, as well as the sources or organizations behind the technologies~\cite{huang2013don,kramer1999trust}. While these heuristic-based judgments are indispensable for people to navigate complex technological and social environments, they also lead to biases and errors if inappropriately applied~\cite{kahneman1982judgment}, calling for careful study of inferences people make based on ST features and their potential effect.

Our concept of social transparency in AI systems is informed by the aforementioned perspectives, but with several key distinctions: at the center of our work is a desire to support the explainability of AI systems, particularly in AI-mediated decision-making. We are not merely interested in making \textit{others'} activities visible, but more importantly, how \textit{others' interactions with AI} impact the explainability of the system. Within the view of a human-AI assemblage, in which both AI and people have decision-making agency, it is possible to borrow ideas and interpretative lenses from work studying ST in human-human interactions. To study the effects of ST in AI systems, our first-order focus is on users' sense-making of an AI system and their decision-making process, though it may inevitably impact their organizational behaviors as well.

\section{Social Transparency in AI systems: a scenario based design exploration}
\label{4w}
After identifying an epistemic blind spot of XAI, we propose adding Social Transparency (ST) into AI systems--incorporating social-organizational contexts to facilitate explainability of AI’s recommendations. This definition is intentionally left broad, as we follow a broad definition of explainability--ability to answer the why-question. We borrow the term ST from Stuart et al.~\cite{stuart2012social}, and similarly emphasize both making visible of other people in the human-AI assemblage, and other people’s interactions with the “source”, in our case, the AI system.  Different from Stuart et al., which proposed the ST concept retrospectively at a time when ST enabling features were pervasive in CSCW systems, we had to consider, prospectively, what kind of features to add to an AI system to make ST possible. 

As a formative step, our goal was not to develop a finished treatise of ST in AI systems. Rather, we intended to create an exemplary design of an AI system with ST and use it to conduct formative studies to advance our conceptual development.  We opted for a scenario-based design (SBD) method. SBD suspends the needs to define system operations by using narrative descriptions of how a user uses a system to accomplish a task~\cite{rosson2009scenario}. SBD allows interpretive flexibility in a user journey by balancing between roughness and concreteness. SBD is an appropriate choice for our investigation because it is a method oriented for “envisioning future use possibilities” ~\cite{rosson2009scenario},  focusing on people’s needs, evocative, and has been adopted in prior XAI design work~\cite{wolf2019evaluating}.

We started with a range of AI-mediated decision-making scenarios around cybersecurity, hiring (employment), healthcare, and sales, where a user encounters an AI recommendation and seeks answer to a \textit{why}-questions about the recommendation, e.g. “why should I accept or trust the recommendation”. We ran 4 workshops with a total of 21 people from 8 technology companies who are users or stakeholders of relevant AI systems. The scenarios started in a textual form, then we engaged participants in drawing exercises to create  visual mock-ups of these scenarios (hereby referred to as visual scenarios), and brainstorming together what kind of information they wanted to see about \textit{other users} of the AI system, and \textit{other users’ interactions with the AI system} if they were the user. When it came to types of design feature that could encode relevant socio-organizational context, people had many suggestions. For instance, suggestions about  knowing what happened to other people getting recommendations from the AI systems, who got the recommendations, etc. quickly emerged in the discussions. The ideas converged to what our participants coined as the “4W”—\textit{who} did \textit{what} with the AI system, \textit{when}, and \textit{why} they did what they did— in order to have adequate socio-organizational context around the AI-mediated decisions.

\begin{figure}[tbh]
    \caption{Visual scenario used in the interviews, labeled by blocks to be revealed in the interview in order: (1) Decision information and model explanation: Information of the current sales decision, the AI's recommended price and a ``feature importance'' explanation justifying the model's recommendation, inspired by real-world pricing tools; (2) ST summary: Beginning of ST giving a high-level summary of how many teammates in the past had received the recommendation and how many sold at the recommended price; (3-5): ST blocks with "4W" features containing the historical decision trajectory of three other users.}
    \centering    
    \includegraphics[width=14.95cm]{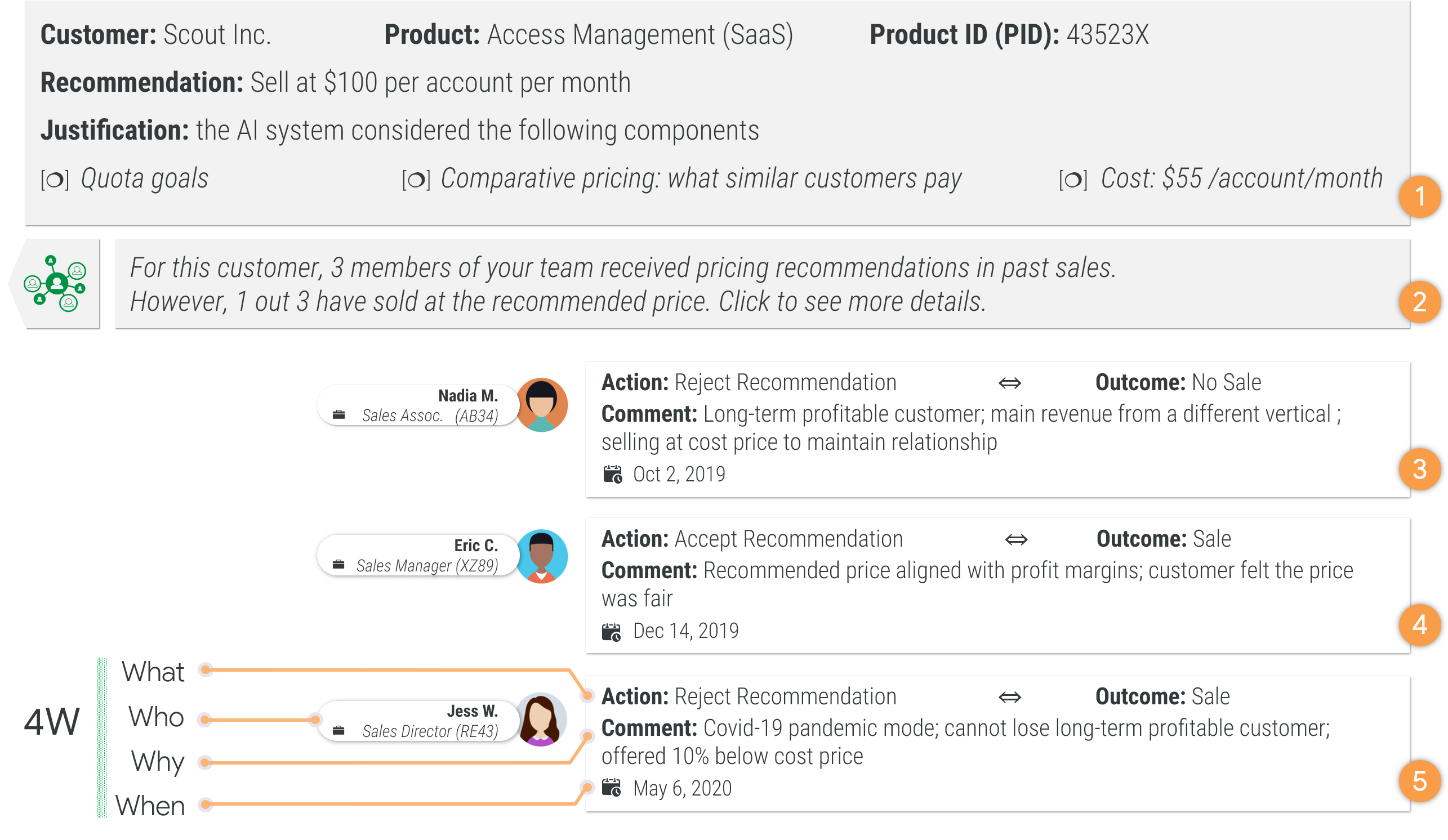}
    \label{fig:visual scenario}
    \Description[Figure showing the anatomy of the visual scenario (sales context) we explored in the interviews.]{{Visual scenario used in the interviews, by labeled blocks: (1) Decision information and model explanation: Information of the current sales decision, the AI's recommended price and a ``feature importance'' explanation justifying the model's recommendation, inspired by real-world pricing tools; (2) ST summary: Beginning of ST giving a high-level summary of how many teammates in the past had received the recommendation and how many sold at the recommended price; (3-5): ST blocks with "4W" features containing the historical decision trajectory of from each of the colleagues.}}
    
\end{figure}

We note an interesting observation that the 4W share similarity with the design elements for group awareness in groupware work~\cite{gutwin2002descriptive}, with the exception of “why”, which is core to  explainability. When thinking how to represent the “why”, participants suggested an open ended textual representation to capture  the nuances behind a decision. Eventually, we settled on a design of a “commenting” feature (why) together with traces of others’ interactions with the AI system’s recommendations (what), their identities (who) and time of interactions (when). In the rest of the paper, we refer to these constitutive design elements of ST as 4W. Figure~\ref{fig:visual scenario} shows the final visual scenario with the 4W features used in the interview study. 

We chose a sales scenario around an AI-mediated price recommendation tool, since it appeared to have a broader reach and accessibility even for workshop participants who did not work in a sales domain. In the study, we intended to interview sellers as targeted users of such an AI system, and also non-sellers to explore the transferability of the ST concept to other AI domains, as we will discuss in detail in the next section.

\textit{Design choices in the visual scenario:} We ran 4 pilot studies to finalize the design of the visual scenario in Figure~\ref{fig:visual scenario}, and the procedure to engage participants with the design. We scoped the number of 4W blocks to three to strike a balance between a variety of ST information and avoiding overwhelming the participants, based on what we learned from the pilot studies. Each of the 4W are represented by one or more design features: accepting and rejecting the AI (action [what]), succeeding and failing to make the sale (outcome [what]), one's name, profile picture and organizational role ([who]), a comment on the reasons behind the action ([why]), and a time stamp ([when]). Contents in these components were inspired by the workshop discussions, and showcase a range of socio-organizational contexts relevant to the decision. The pilot runs revealed that presenting the entire visual scenario creates cognitive and visual clutter. Therefore, for the interview, we decided to reveal the five blocks shown in Figure~\ref{fig:visual scenario} one by one, with the interviewer verbally presenting the narrative around each block.

\section{Study Methods}

In this section we share the methodological details of the semi-structured interviews. 

\subsection{Recruitment}
As mentioned, we intended to recruit both sellers and non-sellers, who are stakeholders of other AI-mediated decision-making domains. Stakeholders are not limited to end users. We also welcomed different perspectives from designers, data scientists, etc. With this in mind, we recruited participants from six different companies, including a large international technology company where we were able to recruit from multiple lines of products or sales divisions. The recruitment was initiated with an online advertisement posted in company-wide group-chat channels that we considered relevant, followed up by snowball sampling. The advertisement stated two recruiting criteria: First, they needed to have direct experience using or developing or designing an AI system.  Second, the AI system should be interacted by multiple users, preferably with multi-user decision-making. We verified that these criteria were met through a series of correspondence (via online messaging) where each participant shared samples of the AI system they intended to discuss.  

A total of 29 participants were recruited (17 self-identified as females while the rest as males). The recruitment of sellers turned out to be challenging, given their very limited availability. By using snowball sampling, we were able to recruit 8 sellers. For non-sellers, the snowball sampling resulted in participants clustered in two major domains  – healthcare and cybersecurity. We conducted the study in the middle of Covid-19 (a global pandemic in 2020), which added non-trivial burden to the recruitment process and limited our interviews to a remote setting using video conferencing tools. Participants’ ID, role, domains and domain experience is shared in Table~\ref{table:participant details}. To facilitate traceability in the data presented hereafter, we differentiate sellers and non-sellers by appending the participant ID with \textit{-S} for sellers and \textit{-NS} for non-sellers (e.g., 1-S for a seller and 2-NS for a non-seller).

\subsection{Interview procedure}

The semi-structured interviews were conducted online with screen-sharing for the interviewer to present the visual scenario. All interviews were video recorded including the screen activities. The interview had 4 main parts. In the \textit{first} part, after gaining informed consent, we asked participants to share about an AI system that they were currently engaged with, focusing on their or their users’ needs for explainability. We also inquired about the socio-organizational context around the use case, both before and after the AI system was introduced.

The \textit{second} part involved a deep dive into the speculative design with a walk-through of the visual scenario in Figure~\ref{fig:visual scenario}. This is where we explored how incorporation of ST can impact an AI-mediated decision-making scenario, as we revealed the different blocks of the visual scenario in a sequenced manner. Participants were asked to play the role of a salesperson trying to pitch a good price for an Access Management software to Scout Inc. (a client). In the first block revealed, the AI not only recommends a price, but also shows  a technical explanation--a set of model features (e.g., cost price, quota goals, etc.) justifying the recommendation. Once the participant showed a good enough understanding on the Decision Information and Model Explanation portion (block 1 in Figure~\ref{fig:visual scenario}), we asked the participant to give a price they would offer and their confidence level (between 1-10, 10 being extremely confident) given what they saw on the screen. Next, we revealed the social transparency portions. First, it was the ST Summary (block 2 in Figure~\ref{fig:visual scenario}) followed by each of the 4W blocks (block 3-5 in Figure~\ref{fig:visual scenario}). We allowed participants to read through the content and guided them through any misunderstandings. They were encouraged to think-aloud during the whole process. Following this, we asked participants to share the top three reactions to the addition of the ST features, either positive or critical. After that, participants were asked to share their final price and confidence level. In addition, we asked them to rank the importance of the 4W (\textit{who}, \textit{what}, \textit{when}, and \textit{why}) for their decision-making process and justify their ranking.

The \textit{third }part was about zooming out from the visual scenario and brainstorming plausible and impactful transfer scenarios of ST in domains our participants resided. At this point, we also gave them a conceptual definition and some vocabulary around ST so that they could brainstorm with us effectively. The goal of this part was to explore the design and conceptual space of ST in domains beyond the sales scenario. For sellers, this meant transferring to their own sales work context, which helped refining our own understanding of the sales scenario. Once participants shared their thoughts on the transferability of ST, they ranked the 4W in the transfer use cases. We wanted to see if there are variations in the rankings as the context switches—an aspect we discuss in the Findings section. 

The \textit{fourth} and final part involved discussions around potential unwanted or negative consequences of ST as well as reflective conversations on how incorporation of ST can impact explainability of AI systems.

In summary, in addition to open-ended discussions, our interview collected the following data points from each participant: original and updated price decisions and associated confidence levels, rankings of 4W for both the sales scenario and one's own domain. While our study was not designed to quantitatively evaluate the effect of ST, we will report summary statistics of these data points in the Findings section, which helped guiding our qualitative analysis. 

\subsection{Qualitative Analysis of the interviews}
The interviews lasted 58 minutes on average. We analyzed the transcription of roughly 29 hours of interview data using a combination of thematic analysis\cite{braun2006using} and grounded theory~\cite{strauss1994grounded}. Using an open coding scheme, two authors independently went through the videos and transcription to produce in-vivo codes (directly from the data itself). Then we separately performed a thematic analysis, clustering the codes from in-vivo coding to themes. We iteratively discussed and agreed upon the codes and themes, constantly comparing and contrasting the topics each of us found, refining and reducing the variations in each round till consensus was reached. We grouped the codes and themes at the topic level using a combination of mind-mapping and affinity diagramming. Our results section below is organized thematically, with the top-level topics as subsections. When discussing each topic, we highlight codes that add to that topic in \textbf{bold}.

\begin{table*}
\centering
\caption{Participant details}
\label{table:participant details}
\begin{tabular}{llll} 
\toprule
\multicolumn{1}{l}{Participant ID} & Role               & Domain             & \multicolumn{1}{l}{Years of Experience~}  \\ 
\midrule
1-S                                 & Seller             & Sales              &  > 10                                       \\
2-NS                                  & Designer         & Cybersecurity      &  > 5                                        \\
3-NS                                  & Designer           & Finance and Travel &  > 5                                        \\
4-NS                                  & Consultant         & Gov and Non-profit &  > 3                                        \\
5-S                                  & Seller             & Sales              &  > 5                                        \\
6-NS                                  & Designer           & Health- Oncology   &  > 5                                        \\
7-NS                                  & Data Scientist     & Cybersecurity      &  > 8                                        \\
8-S                                  & Seller             & Sales              &  > 3                                        \\
9-NS                                  & Designer           & Health- Radiology  &  > 5                                        \\
10-NS                                 & Data Scientist     & Cybersecurity      &  > 10                                       \\
11-NS                                 & Designer           & Health             &  > 3                                        \\
12-NS                                 & Designer          & Cybersecurity      &  > 5                                        \\
13-S                                 & Seller             & Data Analytics     &  > 10                                       \\
14-NS                                 & Data Scientist     & NLP                &  > 5                                        \\
15-NS                                 & Designer           & Health- Radiology  &  > 5                                        \\
16-NS                                 & MD/ Data Scientist & Health- Oncology   &  > 10                                       \\
17-NS                                 & Manager            & HR                 &  > 5                                        \\
18-S                                 & Seller             & Sales              &  > 3                                        \\
19-S                                 & Seller             & Sales              &  > 3                                        \\
20-S                                 & Seller             & Sales              &  > 10                                       \\
21-S                                 & Seller             & Sales              &  > 3                                        \\
22-NS                                 & SOC analyst        & Cybersecurity      &  > 3                                        \\
23-S                                 & Seller             & Sales              &  > 5                                        \\
24-NS                                 & SOC analyst        & Cybersecurity      &  > 5                                        \\
25-NS                                 & SOC analyst        & Cybersecurity      &  > 3                                        \\
26-NS                                 & SOC analyst        & Cybersecurity      &  > 5                                        \\
27-NS                                 & SOC Data Scientist & Cybersecurity      &  > 5                                        \\
28-NS                                 & SOC Architect      & Cybersecurity      &  > 10                                       \\
29-NS                                 & SOC analyst        & Cybersecurity      &  > 5                                        \\
\bottomrule
\end{tabular}
\Description[Table showing participant details]{In this table, we share the participant details. From left to right, the columns are Participant ID, Role, Domain, and Years of Experience}
\end{table*}
\section{Findings}
We begin by sharing how participants' own experience with AI systems demonstrates that technical transparency alone does not meet their explainability needs. They need context beyond the limits of the algorithm. Next, based on how participants reacted to the incorporation of ST in the design scenario, we unpack what context could be made visible by ST and break down the implications at three levels: \textbf{technological (AI)}, \textbf{decision-making}, and \textbf{organizational}, as summarized in Table~\ref{tab:conceptual framework}. We further discuss specific aspects of socio-organizational context that the 4W design features carry and their effects, summarized in Table~\ref{tab:4W}. Based on input from non-seller participants, we also share insights about the potential transferability of ST beyond the sales domain. We end this section with participants’ discussions on the challenges, risks, and tensions of introducing ST into AI systems.
 
As mentioned, all participants, both sellers and non-sellers, experienced the sales scenario and reflected on its transferability to their own domains. Our analysis revealed substantial alignment between the two groups, possibly due to the accessible nature of our intentional choice of a sales domain and the content in the scenario. With the exception of Section~\ref{transferability}, which focuses on non-sellers' reflection on transferability of ST, we report the results combining the two groups, but mark their IDs differently (-\textit{S} or -\textit{NS}) as shwon in Table~\ref{table:participant details}. 

\subsection{Technical Transparency is not enough}
As we began each interview with participants' own experience with AI systems, a core theme that lies at the heart of our findings is the realization that solely relying on technical or algorithmic transparency is not enough to empower complex decision-making. There is a shared understanding that AI algorithms cannot take into account all the contextual factors that matter for a decision: “not everything that you need to actually make the right decision for the client and the company is found in the data” (P25-NS). Participants pointed to the fact that even with an accurate and algorithmically sound recommendation, “there are things [they] never expect a machine to know [such as] clients’ allegiances or internal projects impacting budget behavior” (P1-S).  Often, the context of social dynamics that an algorithm is unable to capture is the key: “real life is more than numbers, especially when you think of relationships” (P12-NS). Discussing challenges in interpreting and using AI recommendations in Security Operation Centers (SOC), P29-NS highlighted the need for awareness of others’ activities in the organizational context: 
\begin{quote}
    Sometimes, even with perfect AI, the most secure thing is to do nothing because you don’t know what the machine doesn't know. There is no centralized process to tell us the context of what's going on elsewhere, what others are doing. One move has ripple effects, you know. So instead of using [the AI’s recommendation], they end up basically doing the most secure thing-- don't touch anything. That’s where the context helps from your colleagues. That’s how actually work really gets done. (P29-NS, a SOC director)
\end{quote}

Moreover, even when provided, technical transparency is not always understandable for end users. While describing how he uses an AI-assisted  pricing tool, this seller pointed to how the machine explained itself by sharing a “confidence interval” along with a description of how the AI works, which was meaningless to him:     
\begin{quote}
    I hate how it just gives me a confidence level and gibberish that the engineers will understand. There is zero context. The only reason I am able to use this tool is [through] guidance from other sellers who gave me the background information on the lead I needed to generate a quote worth their time. (P23-S, senior salesperson using a pricing tool to generate a quote)
\end{quote}

In complex organizational settings, answers to the why-question, i.e. knowledge needed to understand and take informed action for an AI mediated decision, might lie outside the bounds of the machine. As highlighted above, participants repeatedly desired for ``context'' to ``fill in the gaps'' (P27-NS). The ST information in our design scenario is intended to provide such context. After going through the ST portion, 26 out of the 29 participants lowered their sales prices, resulting in a mean final price of $\$73.8$ (SD=$\$15.8$), compared to a mean initial price of $\$110.7$ (SD=$\$57.2$) based only on the AI's recommended price of $\$100$. 24 out of the 29 participants also increased the confidence ratings for their decisions, resulting in a mean final confidence score of $8.3$ out of 10 (SD=$0.9$), compared to a mean initial confidence score of $6.4$ (SD=1.7). These patterns suggest that ST information helped participants to set their price more cautiously and feel more confident about their decisions, by  “help[ing] [them] to understand the situation more holistically” (P19-S). This participant succinctly summarized this perspective at the end of her interview: 
\begin{quote}
    You can’t just get everything in the data that trains the model. The world doesn’t run that way. So why rely just on the machine to make sense of things that are beyond it? To get a holistic sense of the "why" you should or should not trust the AI, you need context. So the context from Social Transparency adds the missing piece to the puzzle of AI explainability”. (P2-NS, a designer of cybersecurity AI systems)
\end{quote}

Now we analyze participants' reaction and reflection from seeing ST features, and unpack the “context” made visible by ST and its effects at three levels: \textbf{technological (AI)}, \textbf{decision-making}, and \textbf{organizational}. For the subsections below each dedicated to a level of context, we begin by summarizing the effects of ST, with codes from the data in bold. These results are summarized in Table~\ref{tab:conceptual framework}.

\begin{table*}
\centering
\caption{Results on the three levels of context made visible by ST and their effects. ``--'' in the last column indicates first-order to second-order effect(s)}
\label{tab:conceptual framework}
\resizebox{\linewidth}{!}{%
\begin{tabular}{>{\hspace{0pt}}p{0.152\linewidth}>{\hspace{0pt}}p{0.3\linewidth}>{\hspace{0pt}}p{0.479\linewidth}} 
\toprule
\textbf{Levels}    & \textbf{Context made visible}                                                                                                                        & \textbf{Effects of the visibility}                                                                                                                                                                                    \\ 
\midrule
Technological (AI) & Trajectory of AI’s past decision outputs and people’s interactions with these outputs  & Tracking AI performance  --~ Calibrate AI trust~ ~ ~ \par{} Infusing human elements  --~ Calibrate AI trust~                                                                                 \\ 
\midrule
Decision-making    & Local context of past decisions and in-situ  access to decision-related (crew) knowledge                                                     & Actionable insights -- Improve decisions; Boost decision~ confidence; Support follow-up actions \par{} Social validation~ -- Decision-making resilience;~ ~ ~AI contestability\textit{~}  \\ 
\midrule
Organizational     &  Organizational meta-knowledge and practices                                                            &
Understanding organizational norms and values -- Improve decisions; Set job expectation\par
Fostering auditability and accountability\par Expertise location -- Develop TMS                       \\
\bottomrule
\end{tabular}
}
\Description[Table showing implications of ST.]{This table shows the results on the three levels of context made visible by ST and their effects. ``--'' in the last column indicates first-order to second-order effect(s)}
\end{table*}

\subsection{Technological (AI) context made visible} 
ST makes visible the socially-situated \textbf{technological context}: the trajectory of AI’s past decision outputs as well as people’s interactions with these technological outputs. Such contextual information could help people \textbf{calibrate trust in AI}, not only through \textbf{tracking AI performance}, but also by \textbf{infusing human elements} in AI that could invoke social-based perception and heuristics. 

Records of others’ past interactions with the AI system paints a concrete picture of the AI performance, which technical XAI solutions such as performance metrics or model internals would not be able to communicate. Participants felt that the technological context they understood through ST helped them better gauge the AI’s limitations or “actual performance of the AI” (P10-NS). In fact, after going through the sales scenario, many reported on re-calibrating their trust in the AI, which is key to preventing both over-reliance of AI and “AI aversion”~\cite{dietvorst2015algorithm}: 
\begin{quote}
    Knowing the past context helps me understand that the AI wasn’t perfect. It’s almost like a reality check. The comments helped because real life is more than numbers. I am more confident in myself that I am making the right decision but less trust[ing] the AI. (P12-NS, an XAI designer)
\end{quote}

ST could also affect people’s perception of and trust in the AI system by infusing the much needed human elements of decision-making in the machine. Participants from each of the domains (sales, cybersecurity, healthcare) highlighted that ``there is a human aspect to [their] practice'' (P6-NS), something that ``can never be replaced by a machine'' (P6-NS). Adding these human elements allows one to apply familiar social rules. Many participants commented on a “transitive trust” (P4-NS) from trusting their peers -- ``people are trained to believe [their] peers and trust them'' (P25-NS) -- to trusting the AI system, if others were using the AI systems or accepting the AI’s recommendations. For instance, in the sales domain of the scenario, a transitive trust is often fostered by an organizational hierarchy or job seniority ``as precedence and permission for doing the right thing’’ (P12-NS). Radiologists often want to “know who else used the same logic and for what reason” (P15-NS) when working with AI-powered diagnostic tools. In cybersecurity, “knowing that a senior analyst took a certain route with the recommendation [can be] the difference maker” (P28-NS). Some participants also commented on a positively perceived “humanizing effect” of AI by adding ST, that “[users] would potentially adopt... showing them like [AI is] supporting you not replacing you'' (P6-NS).

 The above discussions show that ST could support forming appropriate trust and evaluation of AI through two essential routes, as established in prior work on trust and credibility judgment of technologies~\cite{metzger2010social,metzger2013credibility,sundar2008main}: a central route that is based on a better understanding of the AI system, and a peripheral route by applying social or group-based heuristics such as social endorsement, authority, identity, or social presence heuristics. While the central route tends to be cognitively demanding, the peripheral route is fast and easy, and could be especially impactful to help new users to enhance their trust and adoption of an AI system. 

\subsection{Decision-making context made visible}

ST also makes visible the \textbf{decision context} -- the local context of past decisions -- for which many participants described as “in-situ access” to “crew knowledge”
\footnote{The original term used by most participants was "tribal knowledge", which is a term often used in business and management science to refer to unwritten knowledge within a company. We note that, from an Indigenous perspective particularly in North America, the words "tribe" and "tribal" connote both an official status as a recognized Nation, and also a profound sense of identity, often rooted in cultural heritage, a specific ancestral place, and a lived experience of the on-going presence of tribal elders and ancestors (past, present, and future). In our case, participants used the word "tribal" in a non-Indigenous meaning. Being sensitive to potential mis-use of the word, we engaged in critical conversations with potentially affected community members to understand their perspectives. The conversations revealed that it is best to avoid using that word. We went back to the participants who used the word "tribal" and asked if "crew" captures the essence of what they meant by "tribe". All of them agreed that the words were interchangeable. As such, we only present the data using the term "crew knowledge". }.
We will first elaborate on the notion of \textbf{crew knowledge}, then discuss how a combination of decision trajectory, historical context and elements of crew knowledge could (1) lead to \textbf{actionable insights}, which could \textbf{improve decision-making}, \textbf{boost decision confidence} and \textbf{support follow-up actions}; (2) provide social validation that facilitates \textbf{decision-making resilience} and \textbf{contestability of AI}.

The notion of \textit{crew knowledge} emerged during our discussions with many participants regardless of their domains. When asked to elaborate on the concept, participants defined it as “informal knowledge acquired over time through hands-on experience”, knowledge that is not typically “gained through formal means, but knowledge that’s essential to do the job” (P8-S). Crew knowledge is learned "via informal means, mainly through colleague interactions” (P23-S). It can encode “idiosyncrasies like client specific quirks” (P27-NS). Participants referred to their team as their “crew", with a sense of identity and belonging to a community membership. We can think of crew knowledge as informal or tacit knowledge that is acquired over time and locally-situated in a tight-knit community of practice--an aggregated set of “know-hows” of sorts. While ST features may not explicitly encode a complete set of crew knowledge, they provide in-situ access to the vital context of past decisions that carry elements of crew knowledge.

The central position of crew knowledge in participants’ responses demonstrates that ST can act as a vehicle for knowledge sharing and social learning  in “one consolidated platform” (P21-S).  Participants repeatedly mentioned two types of insights they gained from ST to be particularly actionable for AI-mediated decisions. The first is additional variables important for the decision-making task that are not captured in the AI’s feature space. For example: “I have a lot more variables that I'm aware of to consider, like, the whole pandemic thing...”(P12-NS). These additional variables are often tacit knowledge, idiosyncratic to the decision, or constantly changing, making them impossible to be formalized in an algorithm. ST could support in-situ access to these variables.  

Second, ST supports analogical reasoning with similar decisions and their actual outcomes. Participants exhibited a tendency to reason about the similarity and differences between the contexts of the current decision and past decisions made visible by ST. For example: “what did other oncologists do for a patient like that? So, what treatments were chosen for patients like this person?” (P6-NS)  or “I see the reasoning why they didn't pay the recommended price the other time... but those were different circumstances and look now, they're were growing customer and we need to push them up closer to the more profitable price” (P1-S).

Gaining actionable insights could ultimately boost decision confidence, as most participants commented on increasing their confidence in the final price. We also observed an interesting bifurcation on how they conceptualize confidence in the AI versus confidence in oneself after being empowered with knowledge about the decision context. This quote encapsulated that perspective well: 

\begin{quote}
    The system will go by the numbers but I have my “instincts” thanks to my [crew] knowledge. With these comments, you can say I also have my team’s “instincts” to help me. So I am less confident on the AI but more in myself due to the 360 view I have of things-- I have more information than the machine. (P22-NS)
\end{quote}

Moreover, participants commented that learning from the decision context could also support follow-up actions such as interacting with clients or “justifying” (P1-S)  the decision to supervisors, as illustrated in the quote below:  
\begin{quote}
    And I actually learned a lot. I learned from their comments... I feel like this is an education for the next sale. Even [if it is] another customer, I will be more confident...and know what to do with [the AI's recommendation] because I know how to evaluate it. (P12-NS)
\end{quote}

Learning about past decisions from others, especially higher echelons of the organizational hierarchy, also provided social validation. Social validation can reduce the feelings of individual vulnerability in the decision-making process. While going through the sales scenario, participants would often comment how “the director (Jess) offering discounts gives [them] the permission to do the same” (P12-NS). Being able to have a “direct line of sight into the trajectory of how and why decisions were done in the past” (P24-NS) can make one feel empowered, especially if one has to contest the AI. For most participants, their use of AI systems was mandated by their employers. Many a time, the technology got in the way, becoming a “nuisance” (P1-S) they needed to “fight” (P5-S). Contesting the machine often requires time-consuming reporting and manual review, which creates a feeling that one “can’t just say no to the AI” (P25-NS). This participant elaborated on the vulnerability and how social validation could empower one to act: 
\begin{quote}
    People are afraid—they don’t want to screw up. You look like a dumb*** if you end up in the war room and say you goofed up because you blindly followed the machine. Even if you have at least one other person doing something similar with the AI, you are safe. Just that knowledge is enough to act less scared. [If] your neck is on the line, someone else’s is also on the line. It distributes the risk. (P26-NS)
\end{quote}

\subsection{Organizational context made visible}
Lastly, ST gives visibility to the broader \textbf{organizational context}, including the meta-knowledge about the organization such as who knows what and organizational practices. Different from decision context, which makes visible knowledge localized to the decision, organizational context reflects macro-information about the organization. This differentiation shares similarity with the concepts of content versus interaction transparency in Stuart et al.'s ST in social network ~\cite{stuart2012social}, which emphasizes that transparency of others' interactions enables awareness of ``normative behaviors as well as the social structure within a community''. We observed that such awareness could then: (1) inform an \textbf{understanding of organizational norms and values} that help \textbf{improve decision-making} and \textbf{calibrate people’s overall job expectations}; (2) foster \textbf{accountability and auditability}, and 3) facilitate \textbf{expertise location}, and if done right, over time the \textbf{formation of a Transactive Memory System (TMS)}~\cite{yoo2001developments}. In short, organizational context made visible by ST could foster effective collective actions in the organization and strengthen the human-AI assemblage.

Visibility of others' actions in an organization (what’s done) could translate into an understanding of organizational norms (what’s acceptable) and values (what’s important), which might be otherwise neglected since "norms are often not enshrined in a rule book" (P25-NS).  From the comments in the scenario, participants were informed of organizational norms: “the fact that a director offered the discount below cost price means that this is something that's acceptable. I might be able to do” (P12-NS) and values: “seeing Jess [the director in our scenario] give such a steep discount and noting how she did it to retain a customer, tells [us] that relationship matters to this company” (P28-NS). This type of insight is crucial for making informed decisions and setting overall job expectation, especially for new employees to “learn about the culture of the company” (P16-NS). The following participant succinctly summarized this point: 
\begin{quote}
    The comments...get me a sense of what should be done, what’s expected of me, and what I can also get away with. It tells me what this company values. This helps me understand why certain things are done the way they are, especially if they go against what the AI wanted me to do. This actually explains why I need to do something. (P25-NS).
\end{quote}

The enactment of ST in an AI system shared across an organization enables accountability. Participants felt that if they knew “who did what and why, [then] it provides a nice way to promote accountable actions” (P26-NS). Participants noted that currently there is a level of opaqueness in workers' decision-making processes, making it difficult to uphold accountability, be it during bank audits, sales audits, or standardization on health interventions. ST, according to them, can provide ``peripheral vision'' (P29-NS) that can boost accountability by not only making past decisions traceable, but also socially-situated to better evaluate and attribute responsibilities for, as highlighted by this quote: 
\begin{quote}
    I think these comments would be extremely important for audits and postmortems after an attack. The traceability is huge. (P26-NS, a senior SOC analyst)
\end{quote}

That being said, there is a potential double-edged-sword nature to traceability and accountability, where people might feel they are being watched or surveilled. The same participant (P26-NS) articulated this concern:
\begin{quote}
    You know, there is a dark side to this. If you are part of organizations that love to surveil people, then you are out of luck. That is why organizational culture is so important...
    [In our company], we focus on the problem not the person. But you can’t really say this applies [everywhere].
    (P26-NS)
\end{quote}

ST also provides awareness of organizational meta-knowledge~\cite{leonardi2015ambient}, such as who does or knows what, and who knows whom. Many participants reacted to the scenario with reaching out to relevant people made visible through ST: such as “who was driving that sales” (P3-NS), or “reach out to Jeff just because it's the most recent and find out what's going on” (P5-S).  It shows that ST could potentially solve a pain point for larger, distributed organizations by supporting expertise location.

Beyond expertise sharing, some participants commented that knowing whom to reach out to could facilitate the creation of an “institutional memory” (P28-NS), the passing of “legacy knowledge” (P2-NS), and the ability to  “leverage broader resources to lean on” (P8-S). These comments resonate with the core concept of transactive memory systems (TMS)~\cite{moreland2006transactive,brandon2004transactive}, which explains how a group or organization collectively manages the distribution and retrieval of knowledge across different individuals, often through informal networks rather than formal structures~\cite{nardi2002networkers}. TMS could facilitate employee training and benefit new members:
\begin{quote}
    You can’t survive without institutional memory… [but] it’s never written anywhere and is always in the grapevines. Even if some of it could be captured like this [with ST], then that’s a game changer... Training newcomers is hard especially when it comes to getting that “instinct” on the proper way to react to the [security] alerts. Imagine how different training would be if everything was there in one place!” (P28-NS)
\end{quote}

A TMS could also facilitate a peer-to-peer support system that gives employees a sense of community: 

\begin{quote}
    What I really love is the support system you can potentially create over time using ST. This actually reminds of the knowledge repo[sitory] my colleagues and I have set up where we add our nuggets of client specific wisdom which helps others operate better. 
    As you know, we are a virtual team so having this collective support is crucial. (P27-NS, a SOC data scientist)
\end{quote}

Through tight interactions of the community and repeatedly seeing others' decision processes, a TMS can, over time, enable to formation of a collective mind~\cite{weick1993collective,yoo2001developments}--members of a group form a shared cognitive or decision schema and construct their own actions accordingly. Collective mind is associated with enhanced organizational performance and creativity. Interestingly, one participat speculated on how ST can be construed as “mindware”: 

\begin{quote}
    This almost reminds me of a mindware in a team, sort of like a group mind. Currently, we tie our [security] incident reports to a slack channel and that acts as a storage of our collective memories...
    We even have tagged comments, so when you showed me your thing, it reminded me of that. 
    (P25-NS)
\end{quote}

\begin{table*}
\centering
\caption{Summary of the design features, supported effect, and rank of the ``4W'' features}
\label{tab:4W}
\resizebox{\linewidth}{!}{%
\begin{tabular}{>{\hspace{0pt}}p{0.08\linewidth}>{\hspace{0pt}}p{0.3\linewidth}>{\hspace{0pt}}p{0.43\linewidth}>{\hspace{0pt}}p{0.1\linewidth}} 
\toprule
Category & Design Features                                                                                  & Supported Effect                                                                                                                 & Overall Rank  \\ 
\midrule
What     & Action taken on AI\par{}Decision outcome\par{}Summary statement  & Tracking AI performance\par{}Machine contestability (Social validation)                                                          & 1st           \\
\\
Why      & Comments with rationale justifying the decision      & Tracking AI performance\par{}Actionable insights\par{}Understanding organizational norms and values\par{}Social validation\par{} & 2nd           \\
Who      & Name\par{} Organizational role/ job title \par{}Profile picture                                   & Social validation\par{}Transitive trust (Infusing human elements)\par{}Expertise location                            & 3rd           \\ \\
When     & Timing of the decision                                                                            & Temporal relevance (actionable insights)                                                                                         & 4th           \\
\bottomrule
\end{tabular}
}
\Description[Summary of the design features, supported effect, and rank of the ``4W'' features]{This table shows how each of the 4W-- what, why, who, and when-- are conceptualized as design features, what effect they have on the implications of ST, and their overall rank}
\end{table*}

\subsection{Design for ST: the 4W}

With the effects of ST at the three levels in mind, now we discuss how participants reacted to specific design features that are intended to reflect ST. As discussed in Section~\ref{4w}, our co-design exercises informed the choices of constitutive elements of ST: \textit{Who} did \textit{What}, \textit{When}, and \textit{Why}, referred as the 4W. The reader might recall that participants were asked to rank and justify the relative importance of 4W twice during the interview. The first ranking was done in the sales scenario. The second was done in discussing the transferability to participants' own domains. By explicitly inquiring about their thoughts and preferences around the 4W, it helped us understand the effects of each of these design features in facilitating ST and the remaining challenges.

Despite domain-dependent variations, we found overall patterns of preference that are informative. In the sales scenario, first, participants wanted to know “what happened?” (mean rank= 1.90). If the outcome was interesting, then they wanted to delve deeper into the “why” (mean rank= 1.97), followed by “who” (mean rank= 3.03) did it and “when” (mean rank= 3.10).  The relative order of the 4W remained stable when discussing transferability to participant's individual domains. Table~\ref{tab:4W} summarizes the 4W design features and the types of effect they support based on the codes emerged in the interviews. These codes correspond to the effects of the three levels of context shown in Table~\ref{tab:conceptual framework}.

\subsubsection{What:} In our scenario, the “what” is conveyed by two design features – whether (a) a previous person accepted or rejected the AI’s recommendation and (b) whether the sale was successful or not [the outcome]. There was also a summary feature of What appeared at the top (block 2 in Figure~\ref{fig:visual scenario}). Citing that the outcome is the “consequence of [their] decision” (P16-NS), participants felt that it was a must-have element of ST. Participants referred to the “what” as the “snapshot” of all ST information (P5-S, P8-S, P15-NS, P28-NS) which gave them an overview of the AI’s performance and others’ actions, and guided them to decide “do I want to invest more time and dig through” (P15-NS). As the scenario unfolded in the beginning, seeing the summary What feature often invoked a reaction that one should be cautious from over-relying on the AI’s recommendation, but seek further information to make an informed decision. On a more constructive note, especially when thinking through transfer scenarios in radiology and cybersecurity, participants highlighted the need to present the appropriate level of details so that it does cognitively burden the user-- “the outcome should be a TL;DR. The ‘why’ is there if I am interested” (P29-NS).

\subsubsection{Why:} The “Why” information was communicated in free-form comments left by previous users in our scenario. Participants often referred to the “why” as the “context behind the action” and “to understand the human elements of decision-making” (P17-NS). They felt that the “why” could not only help them understand areas that the technology might be lacking, but also “explain the human and the organization” (P28-NS). “Understanding the rationale behind past decisions allows [one] to make similar decisions… [and] gives you an idea of what you should be doing” (P18-S). Prior rationales can also “give [humans] a justification to reject the machine” (P7-NS) by "know[ing] why someone did something similar” (P28-NS). In short, insights into the why can inform AI performance, provide actionable insights and social validation for the decision, as well as facilitate a better understanding of organizational norms and values. Social validation, in particular, can enable contestability of AI. On a constructive note, participants highlighted the need to process or organize the comments to make them consumable: “not all whys are created equal, [and that there is a] need to ensure things are standardized” (P25-NS). There were concerns that if comments are not quality controlled, they might not serve the purpose of shedding context appropriately. Citing “no one wants a lawsuit on their hands” (P26-NS), participants also suggested the need to be vigilant about compliance and legal requirements to ensure private details (e.g., proprietary information) is not revealed. 

\subsubsection{Who: }The “Who” information in our scenario included multiple elements: a previous user’s name, position and a displayed profile picture. Participants engaged with the implications of “who” at multiple levels. For many, ``who'' was the bare minimum that they needed for expertise location – “if [I] knew who to reach out to… [I] could find out the rest of the story” (P5-S). For others,  knowing someone’s  organizational role or level of experience is more important, because “hierarchy matters” (P16-NS) and one’s experience level influences the “degree of trust we can place on other people’s judgement” (P2-NS). Thus the identity information could affect both social validation for one’s own decision and transitive (dis) trust in AI.  On a constructive note,  some reflected on how “the collective `who' matter[ed]” (P6-NS) and there needs to be consistency across personnel for them to make sense of the decision. Some participants raised concerns of the profile picture and the name displayed in our scenario. They felt that these features can lead to biases in weighing different ST information.  Others welcomed the profile information because it “humanizes” the use of AI (P1-S). Here, the domain of the participant appeared to matter – most salespeople welcomed complete visibility; many of the stakeholders from healthcare and government service domains raised concerns. Such perception differences across domains highlight that we need to pay attention to the values in the community of practice as we design these features. 

\subsubsection{When:} The “When” information is expressed by a timestamp. Participants felt that the timing can dictate “if the information is still relevant” (P11-NS), which informs the actionability of context they gain from ST. Knowing the \textit{when} “puts things into perspective” (P16-NS) because it adds “context to the decision and strengthen[s] the why” (P18-S). Timing was particularly useful when participants deliberated on which prior decision they should give more weight to. One comment in the scenario highlighted how Covid-19 (a global pandemic in 2020) influenced the decision-maker’s actions. At the time of the interviews, the world was still going through Covid-19. The “when” “aligned things with a timeline of events and how they transpired” (P21-S).

\subsection{Transferability of ST to other domains}
\label{transferability}
After participants engaged with the sales scenario, we debriefed them on the conceptual idea of adding ST to AI systems. We then asked them to think of transfer scenarios by envisioning how ST might manifest in their own domains or use cases. As Table~\ref{table:participant details} shows, except for 3 people (P4-NS, P11-NS, P17-NS), our participants came from three main domains: sales, cybersecurity, and healthcare (radiology and oncology). Here we give an overview of how participants viewed the potential needs and impact of ST in cybersecurity and healthcare domains.

\subsubsection{Cybersecurity: } Participants working in cybersecurity domain were unanimous in their frustration about the lack of awareness of how their peers make use of AI's recommendations. They saw a rich space where the incorporation of ST could improve their decision-making abilities and provide social validation to foster decision-making resilience. For example, many participants felt that ST would be extremely useful in ticketing systems, where the SOC analyst is tasked with a binary classification deciding if the threat should be escalated or not. Current AI systems have a high rate of false positive alerting a security threat when there is none. This can be stressful for new analysts, as “newcomers [who] always escalates things because they are afraid” (P22-NS). Others pointed out ST can provide insights into organizational practices in the context of compliance regulations. Participants also highlighted ST’s potential to augment a standardized AI with local contexts in different parts of an organization. This would be particularly useful when the AI was trained on a dataset from the Global North but deployed in the Global South:

\begin{quote}
    A lot of the companies operate internationally, right? So one of the things we struggle with is working with international clients whose laws are different. On top of that, the system is trained in North American data. Cyber threats mean different things to different people—what’s harmless to me can breach your system. So yeah, if we can do something like this to augment the AI, I think we can catch threats better in a personalized manner to the client. Also, justifying things would be easier because now you have data from both sides [humans and AI]. (P26-NS) 
\end{quote}

Most participants highlighted how visibility of the crew knowledge would be instrumental to pass on “client specific legacy knowledge” (P2-NS). In fact, many cybersecurity teams have existing tools to track past decisions “beyond the model details” (P26-NS); for instance, one team manually keep a historical timeline of false positive alerts. This knowledge helps them calibrate their decisions because “no clients like the boy who shouts wolf every single time” (P25-NS). However, none of these aspects were integrated. Some even expressed surprise on the similarity after going through our scenario. This participant commented on how integration of their tracking system could facilitate ST to improve decision-making:

\begin{quote}
    It’s not like we don’t have crew knowledge now, you know. But I never really thought about the whole explainability thing from both sides before you showed me this [pointing to the comments in the sales scenario]. Why just have it from the machine? People are black boxes too, you know. Coming at [explainability] from both ends is kind of holistic. I like it. (P29-NS, a SOC analyst)
\end{quote}

Some participants, mainly data scientists, speculated on how one can use the “corpus of social signals” (P10-NS) to feed back into the machine as training data. They wanted to “incorporate the human elements into the machine” (P14-NS) or expert knowledge of “top” analysts back into the AI. They wished the ingestion to not only improve the AI performance, but also to generate socially-situated “holistic explanations”, a point we come back to later in the Discussions section. 

\subsubsection{Healthcare (Radiology and Oncology):} Our participants in the healthcare space mainly work in the imaging decision-support domain. Participants felt that ST has promising transfer potential because it would facilitate peer-review and cross-training opportunities. For radiologists and oncologists, participants highlighted that doctors need “explainability especially when their mental models do not match with the AI[s’ recommendation]” (P16-NS). This is where peer feedback and review for similar AI recommendations can be instrumental.  One person shared a story of how oncologists rely on  “tumor boards” (a meeting made up of specialized doctors to discuss challenging cases). The goal is to decide on the best possible treatment plan for a patient by collaboratively thinking through similar tough cases. This participant equated the tumor board activity to those in the comments, highlighting how the 4W adds a “personal touch” to situate the information amongst “trustworthy peers” (P6-NS). 

Participants also valued the context brought in by ST for multi-stakeholder problems, such as deciding on treatment plans for patients going through therapy. ST can help ensure the plans are personalized because doctors can not only see the AI’s recommendation that’s trained on a standard dataset, but can also “consult or reach out to other doctors [who have] treated similar patients and what were all the surrounding contexts that dictated the treatment plan” (P6-NS). According to them, one of the strengths of ST was that the technical and the socio-organizational layers of decision support were integrated in one place, presented side by side, as highlighted in the following quote:

\begin{quote}
    You need both [social and technical aspects] integrated. Without integration in one place, context switching just takes a lot of time and no one would use it. Just having these things in one place makes all the difference. It’s funny how we actually IM each other to ask what people did with the AI’s alerts. (P27-NS)
\end{quote}

\subsection{Challenges around ST}
\label{challenge}
Providing ST in AI systems is not without its challenges, risks, and tensions. We discuss four themes that emerged from the interviews on the potential negative consequences of ST. Future work should strive to mitigate these problems.

First and foremost, there is a vital tension between transparency and \textbf{privacy}. Similar issues have been discussed in prior work on social transparency in CSCW~\cite{erickson2003social}. Participants were concerned about making themselves visible to others in the organization, especially with job-critical information such as past performance and competing intellect. Some were also worried that individuals could be coerced into sharing such sensitive information. For example, P6-NS commented, based on her experience working with health professionals, that people may be unwilling to disclose detailed information about their work:

\begin{quote}
     I've definitely got on the phone with colleges where they're like, well, you know, not everybody at my practices [are willing to talk about it]... just everybody having access to the outcome probably is not great. Especially if they're not really in a position...and it just becomes like, a point of contention. (P6-NS)
\end{quote}

Some were concerned about revealing personal information. For example, P2-NS reacted by asking: “do I really want this info about me? Who will see it? What can they do with it?” Several suggested to anonymize the \textit{Who} by revealing only general profiles such as position or present the ST information at an aggregated level. 

The second tension is around \textbf{biases} that ST could induce on decision-making. The most prominent concern is on group-thinking, by conforming to the group or the majority’ choices. Other biases could also happen by following eminent individuals such as someone in a “senior position” (P17-NS) or “a friend” (P14-NS).  As discussed in previous sections, ST could invoke social-based heuristics, which could support both decision-making and judgment of AI. However, biases and cognitive heuristics are inevitably coupled, and should be carefully managed. Users in some domains might be more subject to biases from ST than others. For example, P17-NS were hesitant about introducing ST features into the human resource domain, for example for AI assisted hiring: “issues of bias, cherry picking, and groupthink are much more consequential in HR situation” (P17-NS).

A third challenge is regarding \textbf{information overload and consumption} of ST. While the design scenario listed only 3 comments, participants were concerned about how to effectively consume the information if the number of entries increases, and how to locate the most relevant information in them. There was also a tension in integrating ST in one’s decision-making workflow, which was especially prominent in time-sensitive contexts such as clinical decision-support. Some participants suggested avoiding a constant flow of ST and only provide ST where needed, e.g. “[ST] is not the information that always needs to be up in front of their face, but there should be a way to get back to it, especially when you're building confidence in kind of assistance”(P6-NS). Others suggested providing ST in a structured or processed format such as “summarization” (P17-NS) or “providing some statistics” (P5-S). 

Lastly, a tension for the success of ST lies in the \textbf{incentive to contribute}. While there are clear benefits for consumers of ST, it is questionable whether there is enough motivation for people to take the extra effort to contribute, as illustrated by this quote:
\begin{quote}
    One thing that we found interesting is oncology... It's a really hard sell to get them to give feedback into a system because they're so time pressed for their workflow...they're giving you work for free. Like, systems should be doing this for them... but the system can break if they don't participate in that loop. (P6-NS)
\end{quote}

This is a classic problem in CSCW systems~\cite{grudin1988cscw}, which may require both lowering the barriers and cost to contribute, and incentivizing contributions with visible and justifiable benefits.

\section{Discussion \& Implications}
Our results identify the potential effects of ST in AI systems, provide design insights to facilitate ST, and point to potential areas of challenges. In this section, we discuss three high-level implications of introducing ST into AI systems: how ST could enable holistic explainability, how ST could strengthen the Human-AI assemblage, and some technical considerations for realizing ST to move towards a socially-situated XAI paradigm.   

\subsection{Holistic Explainability through ST}
After participants concluded the scenario walk-through, we debriefed them on the concept of ST and the idea of facilitating explainability of AI-mediated decisions with ST. Despite frequently using AI systems and facing explainability related issues in their daily workflows, many participants were initially surprised by the association between socio-organizational contexts and AI explainability. The surprise was met with a pleasant realization when they reflected on the scenario and how it could transfer to their real-world use cases. Perhaps their reaction is not surprising given that the epistemic canvas of XAI has largely been circumscribed around the bounds of the algorithm. The focus has primarily been on “the AI in X-AI instead of the X [eXplainable], which is a shame because it’s the human who matters” (P25-NS). The following sentiment captures this point:
\begin{quote}
    I was taught to think [that] all that mattered [in XAI] was explanations from the model...This is actually the first time I thought of AI explainability from a social perspective, and I am an expert in this space! This goes to show you how much tunnel-visioned we have been. Once you showed me Social transparency…it was clear that organizational signals can definitely help us make sense of the overall system. It’s like we had blinders or something that stopped us from seeing the larger picture. (P27-NS, a SOC data scientist)
\end{quote}

A most common way participants expressed how ST impacted their “ways of answering the why-question” (P28-NS) was how incorporation of the context makes the explainability more “holistic” (P2-NS, P23-S, P6-NS, P27-NS, P29-NS). Acknowledging that “context is king for explanations… [and] there are many ways to answer why” (P27-NS), participants felt that the ST goes “beyond the AI” to provide “peripheral vision” of the organizational context. This, in turn, allows them to answer their why-questions in a holistic manner. For instance, in SOC situations “there is often no single correct answer. There are multiple correct answers” (P2-NS). Since AI systems “don’t produce multitudes of of explanations”, participants acknowledged that “incorporating the social layer into the mix” can expand the ways they view explainability (P28-NS).  Moreover, the humanization of the process can also make the decision explainable to non-primary stakeholders in a way that technical transparency alone cannot achieve. For instance, participants felt that having the 4W can make it easier to justify the decisions to clients and regulators. 

While in this work our focus is on how a holistic explainability through ST could better support decision-makers, we recognize that there are other types of stakeholders and explanation consumers, as well as other types of AI systems, that could benefit from ST. For example, collecting the 4W in the deployment context could help model developers to investigate how the system performs and why it fails, then incorporate the insights gained about the technological, decision and organizational contexts to improve the model. Auditors or regulatory bodies could also leverage 4W information to better assess the model's performance, biases, safety, etc. by understanding its situated impact. The contributors of ST information are not limited to decision-makers. For example, with automated AI systems where there isn't a human decision-maker involved, its explainability could be facilitated by making visible the social contexts of people who are impacted by the AI systems.

\subsection{Making the Human-AI assemblage concrete}
We highlight that a consequential AI system is often situated in complex socio-organizational contexts, where many people interacting with it.  By bringing the human elements of decision-making to the fore-front, ST enables the humans to be explicitly represented, thereby making the Human-AI assemblage concrete. As one participant put it, the socially-situated context can ensure “the human is not forgotten in the mix of things” (P25-NS). In our Findings section, we discussed how organizational meta-knowledge can facilitate formation of Transactive Memory Systems (TMS)~\cite{hollingshead2003potential, moreland2006transactive, wegner1991transactive}, allowing "who knows what" to be explicitly encoded for future retrieval. 
Over time, the "heedful interrelations" enabled by TMS and repeatedly seeing others' decision processes with the AI through ST could possibly enable a shared decision schema in the community, leading to the formation of a collective mind~\cite{weick1993collective, yoo2001developments}. This collective mind is one that includes AI as a critical player. With this conceptualization future work could explore the collective actions and evolution of human-AI assemblages. 

By prioritizing the view of human-AI assemblage over the AI, adding ST to AI systems calls for critical consideration on what information of the humans and whose information is made visible to whom. Prior work on ST in CSCW systems has warned against developing technologies that make it easy to share information without careful consideration on its longer-term second-order effect on the organization and its members~\cite{stuart2012social}. In Section~\ref{challenge} we identified four potential issues of ST as foreseen by our participants, including privacy, biases, information overload and motivation to contribute, all of which could have profound impact on the functioning of the human-AI assemblage. In general, future work implementing ST in AI systems should take socioechnical approaches to developing solutions that are sensitive to the values of stakeholders and ``localized'' to a human-AI assemblage. For example, regarding the privacy issue, participants were sensitive to how much visibility the rest of the organization has to their shared activities and knowledge. When asked how they might envision the boundaries, participants highlighted that one should “let the individual teams decide because every ‘tribe’ is different” (P28-NS).

\subsection{Towards socially-situated XAI}
Using a scenario-based design (SBD) method, we suspended the needs to define system operations and technical details.  Some practical challenges may arise in how to present the 4W to explanation seekers. The first challenge is to handle the quantity of information, especially to fit into the workflow of the users. In addition to utilizing NLP techniques to make the content more consumable, for example by providing memorization or organizing it into facets, it could also be beneficial to give users filtering options, allow them to define ``similarity'' or choose past examples they want to see. Secondly, there needs to be mechanisms in place to validate the quality and applicability of ST information, since ``not all whys are created equal'' (P25-NS). This could be achieved by either applying quality control on the recorded ST information, or through careful design of interfaces to elicit high-quality 4W information from the contributors. Another caveat is that it is common for an AI system to receive model updates or adapt with usage, so its decisions may change over time. In that case, it is necessary to flag the differences of the AI in showing past ST information. Lastly, in certain domains or organizations, it is not advisable or possible to gather all 4W information, sometimes due to the tension with privacy, biases and motivation to contribute, so alternative solutions should be sought, for example by linking past decision trajectories with relevant guidelines or documentation to help users decode the \textit{why} information when it is not directly available. 

Several participants, especially those with a data science background, suggested an interesting area for technical innovation--if “the AI can ingest the social data” (P12-NS) to improve both its performance and its explanations. While recent work has started exploring teaching AI with human rationales~\cite{Ghai2020XAL}, ST could enable acquiring such rationales in real-usage contexts. As suggested by what participants learned from the 4W, the decision and organizational contexts made visible by ST could help the AI to learn additional features and localized rules and constraints, then incorporate them into its future decisions. Moreover, a notable area of XAI work focuses on generating human-consumable and domain-specific machine explanations by learning from how humans explain~\cite{ehsan2019automated,hind2019ted}, which could be a fruitful area to explore when combined with the availability of 4W information.  That being said, it may be desirable to explicitly separate the technical component (to show how the AI arrives at its decision) and the socio-organizational component (as further support for or caution against AI's decision) in the explanations, as participants had strong opinions to be able to “know how and where to place the trust” (P27-NS).

\section{Limitations \& Future Work}

We view our work as the beginning of a broader cross-disciplinary discourse around what explainability entails in AI systems. With this paper, we have taken a formative step by exploring the concept of Social Transparency (ST) in AI systems, particularly focusing on how incorporation of socio-organizational context can impact explainability of the human-AI assemblage. Given this first step, the insights from our work should be viewed accordingly. We acknowledge the limitations that come with using a scenario-based design~\cite{rosson2009scenario}, including the dependency between the scenario and data. The insights should be interpreted as formative instead of evaluative. We acknowledge that we need to do more work in the future to expand the design space and consider other design elements for ST, further unpack the transferability of our insights, especially where this transfer might be inappropriate. We should also investigate how ST impacts user trust over longitudinal use of ST-infused XAI systems. 

Our conception of ST is rooted in and inspired by Phil Agre's notion of Critical Technical Practice~\cite{agre1997computation,agre1997toward} where we identify the dominant assumptions of XAI and critically question the status quo to generate alternative technology that brings previously-marginalized insights into the center. Agre stated that ``at least for the foreseeable future, [a CTP-inspired concept] will require a split identity -- one foot planted in the craft work of design and the other foot planted in the reflexive work of critique.''~\cite{agre1997toward}. As such, ST will, at least for the foreseeable future be a work-in-progress, one that is continuously pushing the boundaries of design and reflexively working on its own blind spots. We have ``planted one foot" in the work of design by identifying a neglected insight--the lack of incorporation of socio-organizational context as a constitutive design element in XAI-- and exploring the design of ST-infused XAI systems. Now, we seek to learn from and with the broader HCI and XAI communities as we ``plant the other foot'' in the self-reflective realm of critique.

\section{Conclusion}
Situating XAI through the lens of a Critical Technical Practice, this work is our attempt to challenge algorithm-centered approaches and the dominant narrative in the field of XAI. Explainability of AI systems inevitably sits at the intersection of technologies and people, both of which are socially-situated. Therefore, an epistemic blind spot in that neglects the ``socio" half of sociotechnical systems would likely render technological solutions ineffective and potentially harmful. This is particularly problematic as AI technologies enter different socio-organizational contexts for consequential decision-making tasks. Our work is both conceptual and practical. Conceptually, we address the epistemic blind spot by introducing and exploring Social Transparency (ST)--the incorporation of socio-organizational context--to enable holistic explainability of AI-mediated decision-making. Practically, we progressively develop the concept and design space of ST through design and empirical research. Specifically, we developed a scenario-based design that embodies the concept of ST in an AI system with four constitutive elements--\textit{who} did \textit{what} with the AI system, \textit{when}, and \textit{why} they did what they did (4W). Using this scenario-based design, we explored the potential effect of ST and design implications through 29 interviews with AI stakeholders. The results refined our conceptual development of ST by discerning three levels of context made visible by ST and their effects: technological, decision, and organizational. Our work also contributes concrete design insights and point to potential challenges of incorporating socio-organizational context into AI systems, with which practitioners and researchers can further explore the design space of ST. By adding formative insights that catalyzes our journey towards a socially-situated XAI paradigm, this work contributes to the discourse of human-centered XAI by expanding the conceptual and design space of XAI. 
\begin{acks}
With our deepest gratitude, we acknowledge the time our participants generously invested in this project. We are grateful to members of the Human-Centered AI Lab at Georgia Tech whose continued input refined the conceptualizations presented here.
We are indebted to Werner Geyer, Michael Hind, Stephanie Houde, David Piorkowski, John Richards, and Yunfeng Zhang from IBM Research AI for their generous feedback and time throughout the duration of this project. 
Special thanks to Intekhab Hossain and Samir Passi for conversations and feedback throughout the years that have constructively added to the notion of Social Transparency. 
This project was partially supported through an internship at IBM Research AI and by the National Science Foundation under Grant No. 1928586.
\end{acks}

\bibliographystyle{ACM-Reference-Format}
\bibliography{sample-base}


\begin{thebibliography}{112}


\ifx \showCODEN    \undefined \def \showCODEN     #1{\unskip}     \fi
\ifx \showDOI      \undefined \def \showDOI       #1{#1}\fi
\ifx \showISBNx    \undefined \def \showISBNx     #1{\unskip}     \fi
\ifx \showISBNxiii \undefined \def \showISBNxiii  #1{\unskip}     \fi
\ifx \showISSN     \undefined \def \showISSN      #1{\unskip}     \fi
\ifx \showLCCN     \undefined \def \showLCCN      #1{\unskip}     \fi
\ifx \shownote     \undefined \def \shownote      #1{#1}          \fi
\ifx \showarticletitle \undefined \def \showarticletitle #1{#1}   \fi
\ifx \showURL      \undefined \def \showURL       {\relax}        \fi
\providecommand\bibfield[2]{#2}
\providecommand\bibinfo[2]{#2}
\providecommand\natexlab[1]{#1}
\providecommand\showeprint[2][]{arXiv:#2}

\bibitem[\protect\citeauthoryear{Abdul, Vermeulen, Wang, Lim, and
  Kankanhalli}{Abdul et~al\mbox{.}}{2018}]%
        {abdul2018trends}
\bibfield{author}{\bibinfo{person}{Ashraf Abdul}, \bibinfo{person}{Jo
  Vermeulen}, \bibinfo{person}{Danding Wang}, \bibinfo{person}{Brian~Y Lim},
  {and} \bibinfo{person}{Mohan Kankanhalli}.} \bibinfo{year}{2018}\natexlab{}.
\newblock \showarticletitle{Trends and trajectories for explainable,
  accountable and intelligible systems: An hci research agenda}. In
  \bibinfo{booktitle}{\emph{Proceedings of the 2018 CHI conference on human
  factors in computing systems}}. \bibinfo{pages}{1--18}.
\newblock


\bibitem[\protect\citeauthoryear{Abdul, von~der Weth, Kankanhalli, and
  Lim}{Abdul et~al\mbox{.}}{2020}]%
        {abdul2020cogam}
\bibfield{author}{\bibinfo{person}{Ashraf Abdul}, \bibinfo{person}{Christian
  von~der Weth}, \bibinfo{person}{Mohan Kankanhalli}, {and}
  \bibinfo{person}{Brian~Y Lim}.} \bibinfo{year}{2020}\natexlab{}.
\newblock \showarticletitle{COGAM: Measuring and Moderating Cognitive Load in
  Machine Learning Model Explanations}. In
  \bibinfo{booktitle}{\emph{Proceedings of the 2020 CHI Conference on Human
  Factors in Computing Systems}}. \bibinfo{pages}{1--14}.
\newblock


\bibitem[\protect\citeauthoryear{Ackerman, Dachtera, Pipek, and Wulf}{Ackerman
  et~al\mbox{.}}{2013}]%
        {ackerman2013sharing}
\bibfield{author}{\bibinfo{person}{Mark~S Ackerman}, \bibinfo{person}{Juri
  Dachtera}, \bibinfo{person}{Volkmar Pipek}, {and} \bibinfo{person}{Volker
  Wulf}.} \bibinfo{year}{2013}\natexlab{}.
\newblock \showarticletitle{Sharing knowledge and expertise: The CSCW view of
  knowledge management}.
\newblock \bibinfo{journal}{\emph{Computer Supported Cooperative Work (CSCW)}}
  \bibinfo{volume}{22}, \bibinfo{number}{4-6} (\bibinfo{year}{2013}),
  \bibinfo{pages}{531--573}.
\newblock


\bibitem[\protect\citeauthoryear{Agre}{Agre}{1997}]%
        {agre1997toward}
\bibfield{author}{\bibinfo{person}{P Agre}.} \bibinfo{year}{1997}\natexlab{}.
\newblock \showarticletitle{Toward a critical technical practice: Lessons
  learned in trying to reform AI in Bowker}.
\newblock \bibinfo{journal}{\emph{G., Star, S., Turner, W., and Gasser, L.,
  eds, Social Science, Technical Systems and Cooperative Work: Beyond the Great
  Divide, Erlbaum}} (\bibinfo{year}{1997}).
\newblock


\bibitem[\protect\citeauthoryear{Agre and Agre}{Agre and Agre}{1997}]%
        {agre1997computation}
\bibfield{author}{\bibinfo{person}{Philip Agre} {and} \bibinfo{person}{Philip~E
  Agre}.} \bibinfo{year}{1997}\natexlab{}.
\newblock \bibinfo{booktitle}{\emph{Computation and human experience}}.
\newblock \bibinfo{publisher}{Cambridge University Press}.
\newblock


\bibitem[\protect\citeauthoryear{Alqaraawi, Schuessler, Wei{\ss}, Costanza, and
  Berthouze}{Alqaraawi et~al\mbox{.}}{2020}]%
        {alqaraawi2020evaluating}
\bibfield{author}{\bibinfo{person}{Ahmed Alqaraawi}, \bibinfo{person}{Martin
  Schuessler}, \bibinfo{person}{Philipp Wei{\ss}}, \bibinfo{person}{Enrico
  Costanza}, {and} \bibinfo{person}{Nadia Berthouze}.}
  \bibinfo{year}{2020}\natexlab{}.
\newblock \showarticletitle{Evaluating saliency map explanations for
  convolutional neural networks: a user study}. In
  \bibinfo{booktitle}{\emph{Proceedings of the 25th International Conference on
  Intelligent User Interfaces}}. \bibinfo{pages}{275--285}.
\newblock


\bibitem[\protect\citeauthoryear{Arrieta, D{\'\i}az-Rodr{\'\i}guez, Del~Ser,
  Bennetot, Tabik, Barbado, Garc{\'\i}a, Gil-L{\'o}pez, Molina, Benjamins,
  et~al\mbox{.}}{Arrieta et~al\mbox{.}}{2020}]%
        {arrieta2020explainable}
\bibfield{author}{\bibinfo{person}{Alejandro~Barredo Arrieta},
  \bibinfo{person}{Natalia D{\'\i}az-Rodr{\'\i}guez}, \bibinfo{person}{Javier
  Del~Ser}, \bibinfo{person}{Adrien Bennetot}, \bibinfo{person}{Siham Tabik},
  \bibinfo{person}{Alberto Barbado}, \bibinfo{person}{Salvador Garc{\'\i}a},
  \bibinfo{person}{Sergio Gil-L{\'o}pez}, \bibinfo{person}{Daniel Molina},
  \bibinfo{person}{Richard Benjamins}, {et~al\mbox{.}}}
  \bibinfo{year}{2020}\natexlab{}.
\newblock \showarticletitle{Explainable Artificial Intelligence (XAI):
  Concepts, taxonomies, opportunities and challenges toward responsible AI}.
\newblock \bibinfo{journal}{\emph{Information Fusion}}  \bibinfo{volume}{58}
  (\bibinfo{year}{2020}), \bibinfo{pages}{82--115}.
\newblock


\bibitem[\protect\citeauthoryear{Arya, Bellamy, Chen, Dhurandhar, Hind,
  Hoffman, Houde, Liao, Luss, Mojsilovi{\'c}, et~al\mbox{.}}{Arya
  et~al\mbox{.}}{2019}]%
        {arya2019one}
\bibfield{author}{\bibinfo{person}{Vijay Arya}, \bibinfo{person}{Rachel~KE
  Bellamy}, \bibinfo{person}{Pin-Yu Chen}, \bibinfo{person}{Amit Dhurandhar},
  \bibinfo{person}{Michael Hind}, \bibinfo{person}{Samuel~C Hoffman},
  \bibinfo{person}{Stephanie Houde}, \bibinfo{person}{Q~Vera Liao},
  \bibinfo{person}{Ronny Luss}, \bibinfo{person}{Aleksandra Mojsilovi{\'c}},
  {et~al\mbox{.}}} \bibinfo{year}{2019}\natexlab{}.
\newblock \showarticletitle{One explanation does not fit all: A toolkit and
  taxonomy of ai explainability techniques}.
\newblock \bibinfo{journal}{\emph{arXiv preprint arXiv:1909.03012}}
  (\bibinfo{year}{2019}).
\newblock


\bibitem[\protect\citeauthoryear{Austin}{Austin}{2003}]%
        {austin2003transactive}
\bibfield{author}{\bibinfo{person}{John~R Austin}.}
  \bibinfo{year}{2003}\natexlab{}.
\newblock \showarticletitle{Transactive memory in organizational groups: the
  effects of content, consensus, specialization, and accuracy on group
  performance.}
\newblock \bibinfo{journal}{\emph{Journal of applied psychology}}
  \bibinfo{volume}{88}, \bibinfo{number}{5} (\bibinfo{year}{2003}),
  \bibinfo{pages}{866}.
\newblock


\bibitem[\protect\citeauthoryear{Brandon and Hollingshead}{Brandon and
  Hollingshead}{2004}]%
        {brandon2004transactive}
\bibfield{author}{\bibinfo{person}{David~P Brandon} {and}
  \bibinfo{person}{Andrea~B Hollingshead}.} \bibinfo{year}{2004}\natexlab{}.
\newblock \showarticletitle{Transactive memory systems in organizations:
  Matching tasks, expertise, and people}.
\newblock \bibinfo{journal}{\emph{Organization science}} \bibinfo{volume}{15},
  \bibinfo{number}{6} (\bibinfo{year}{2004}), \bibinfo{pages}{633--644}.
\newblock


\bibitem[\protect\citeauthoryear{Braun and Clarke}{Braun and Clarke}{2006}]%
        {braun2006using}
\bibfield{author}{\bibinfo{person}{Virginia Braun} {and}
  \bibinfo{person}{Victoria Clarke}.} \bibinfo{year}{2006}\natexlab{}.
\newblock \showarticletitle{Using thematic analysis in psychology}.
\newblock \bibinfo{journal}{\emph{Qualitative research in psychology}}
  \bibinfo{volume}{3}, \bibinfo{number}{2} (\bibinfo{year}{2006}),
  \bibinfo{pages}{77--101}.
\newblock


\bibitem[\protect\citeauthoryear{Bu{\c{c}}inca, Lin, Gajos, and
  Glassman}{Bu{\c{c}}inca et~al\mbox{.}}{2020}]%
        {buccinca2020proxy}
\bibfield{author}{\bibinfo{person}{Zana Bu{\c{c}}inca}, \bibinfo{person}{Phoebe
  Lin}, \bibinfo{person}{Krzysztof~Z Gajos}, {and} \bibinfo{person}{Elena~L
  Glassman}.} \bibinfo{year}{2020}\natexlab{}.
\newblock \showarticletitle{Proxy tasks and subjective measures can be
  misleading in evaluating explainable AI systems}. In
  \bibinfo{booktitle}{\emph{Proceedings of the 25th International Conference on
  Intelligent User Interfaces}}. \bibinfo{pages}{454--464}.
\newblock


\bibitem[\protect\citeauthoryear{Cai, Jongejan, and Holbrook}{Cai
  et~al\mbox{.}}{2019}]%
        {cai2019effects}
\bibfield{author}{\bibinfo{person}{Carrie~J Cai}, \bibinfo{person}{Jonas
  Jongejan}, {and} \bibinfo{person}{Jess Holbrook}.}
  \bibinfo{year}{2019}\natexlab{}.
\newblock \showarticletitle{The effects of example-based explanations in a
  machine learning interface}. In \bibinfo{booktitle}{\emph{Proceedings of the
  24th International Conference on Intelligent User Interfaces}}.
  \bibinfo{pages}{258--262}.
\newblock


\bibitem[\protect\citeauthoryear{Carvalho, Pereira, and Cardoso}{Carvalho
  et~al\mbox{.}}{2019}]%
        {carvalho2019machine}
\bibfield{author}{\bibinfo{person}{Diogo~V Carvalho},
  \bibinfo{person}{Eduardo~M Pereira}, {and} \bibinfo{person}{Jaime~S
  Cardoso}.} \bibinfo{year}{2019}\natexlab{}.
\newblock \showarticletitle{Machine learning interpretability: A survey on
  methods and metrics}.
\newblock \bibinfo{journal}{\emph{Electronics}} \bibinfo{volume}{8},
  \bibinfo{number}{8} (\bibinfo{year}{2019}), \bibinfo{pages}{832}.
\newblock


\bibitem[\protect\citeauthoryear{Cheng, Wang, Zhang, O'Connell, Gray, Harper,
  and Zhu}{Cheng et~al\mbox{.}}{2019}]%
        {cheng2019explaining}
\bibfield{author}{\bibinfo{person}{Hao-Fei Cheng}, \bibinfo{person}{Ruotong
  Wang}, \bibinfo{person}{Zheng Zhang}, \bibinfo{person}{Fiona O'Connell},
  \bibinfo{person}{Terrance Gray}, \bibinfo{person}{F~Maxwell Harper}, {and}
  \bibinfo{person}{Haiyi Zhu}.} \bibinfo{year}{2019}\natexlab{}.
\newblock \showarticletitle{Explaining decision-making algorithms through UI:
  Strategies to help non-expert stakeholders}. In
  \bibinfo{booktitle}{\emph{Proceedings of the 2019 chi conference on human
  factors in computing systems}}. \bibinfo{pages}{1--12}.
\newblock


\bibitem[\protect\citeauthoryear{Cheon and Su}{Cheon and Su}{2016}]%
        {cheon2016integrating}
\bibfield{author}{\bibinfo{person}{EunJeong Cheon} {and}
  \bibinfo{person}{Norman~Makoto Su}.} \bibinfo{year}{2016}\natexlab{}.
\newblock \showarticletitle{Integrating roboticist values into a Value
  Sensitive Design framework for humanoid robots}. In
  \bibinfo{booktitle}{\emph{2016 11th ACM/IEEE International Conference on
  Human-Robot Interaction (HRI)}}. IEEE, \bibinfo{pages}{375--382}.
\newblock


\bibitem[\protect\citeauthoryear{Cheon and Su}{Cheon and Su}{2018}]%
        {cheon2018futuristic}
\bibfield{author}{\bibinfo{person}{EunJeong Cheon} {and}
  \bibinfo{person}{Norman~Makoto Su}.} \bibinfo{year}{2018}\natexlab{}.
\newblock \showarticletitle{Futuristic autobiographies: Weaving participant
  narratives to elicit values around robots}. In
  \bibinfo{booktitle}{\emph{Proceedings of the 2018 ACM/IEEE International
  Conference on Human-Robot Interaction}}. \bibinfo{pages}{388--397}.
\newblock


\bibitem[\protect\citeauthoryear{Dabbish, Stuart, Tsay, and Herbsleb}{Dabbish
  et~al\mbox{.}}{2012}]%
        {dabbish2012social}
\bibfield{author}{\bibinfo{person}{Laura Dabbish}, \bibinfo{person}{Colleen
  Stuart}, \bibinfo{person}{Jason Tsay}, {and} \bibinfo{person}{Jim Herbsleb}.}
  \bibinfo{year}{2012}\natexlab{}.
\newblock \showarticletitle{Social coding in GitHub: transparency and
  collaboration in an open software repository}. In
  \bibinfo{booktitle}{\emph{Proceedings of the ACM 2012 conference on computer
  supported cooperative work}}. \bibinfo{pages}{1277--1286}.
\newblock


\bibitem[\protect\citeauthoryear{Dattner, Chamorro-Premuzic, Buchband, and
  Schettler}{Dattner et~al\mbox{.}}{2019}]%
        {dattner2019hiring}
\bibfield{author}{\bibinfo{person}{Ben Dattner}, \bibinfo{person}{Tomas
  Chamorro-Premuzic}, \bibinfo{person}{Richard Buchband}, {and}
  \bibinfo{person}{Lucinda Schettler}.} \bibinfo{year}{2019}\natexlab{}.
\newblock \showarticletitle{The Legal and Ethical Implications of Using AI in
  Hiring}.
\newblock \bibinfo{journal}{\emph{Harvard Business Review}} (\bibinfo{date}{25
  April} \bibinfo{year}{2019}).
\newblock
\urldef\tempurl%
\url{https://hbr.org/2019/04/the-legal-and-ethical-implications-of-using-ai-in-hiring}
\showURL{%
Retrieved 26-August-2019 from \tempurl}


\bibitem[\protect\citeauthoryear{Dennett}{Dennett}{1989}]%
        {dennett1989intentional}
\bibfield{author}{\bibinfo{person}{Daniel~Clement Dennett}.}
  \bibinfo{year}{1989}\natexlab{}.
\newblock \bibinfo{booktitle}{\emph{The intentional stance}}.
\newblock \bibinfo{publisher}{MIT press}.
\newblock


\bibitem[\protect\citeauthoryear{Dietvorst, Simmons, and Massey}{Dietvorst
  et~al\mbox{.}}{2015}]%
        {dietvorst2015algorithm}
\bibfield{author}{\bibinfo{person}{Berkeley~J Dietvorst},
  \bibinfo{person}{Joseph~P Simmons}, {and} \bibinfo{person}{Cade Massey}.}
  \bibinfo{year}{2015}\natexlab{}.
\newblock \showarticletitle{Algorithm aversion: People erroneously avoid
  algorithms after seeing them err.}
\newblock \bibinfo{journal}{\emph{Journal of Experimental Psychology: General}}
  \bibinfo{volume}{144}, \bibinfo{number}{1} (\bibinfo{year}{2015}),
  \bibinfo{pages}{114}.
\newblock


\bibitem[\protect\citeauthoryear{Dodge, Liao, Zhang, Bellamy, and Dugan}{Dodge
  et~al\mbox{.}}{2019}]%
        {dodge2019explaining}
\bibfield{author}{\bibinfo{person}{Jonathan Dodge}, \bibinfo{person}{Q~Vera
  Liao}, \bibinfo{person}{Yunfeng Zhang}, \bibinfo{person}{Rachel~KE Bellamy},
  {and} \bibinfo{person}{Casey Dugan}.} \bibinfo{year}{2019}\natexlab{}.
\newblock \showarticletitle{Explaining models: an empirical study of how
  explanations impact fairness judgment}. In
  \bibinfo{booktitle}{\emph{Proceedings of the 24th International Conference on
  Intelligent User Interfaces}}. \bibinfo{pages}{275--285}.
\newblock


\bibitem[\protect\citeauthoryear{Doshi-Velez and Kim}{Doshi-Velez and
  Kim}{2017}]%
        {doshi2017towards}
\bibfield{author}{\bibinfo{person}{Finale Doshi-Velez} {and}
  \bibinfo{person}{Been Kim}.} \bibinfo{year}{2017}\natexlab{}.
\newblock \showarticletitle{Towards A Rigorous Science of Interpretable Machine
  Learning}.
\newblock \bibinfo{journal}{\emph{stat}}  \bibinfo{volume}{1050}
  (\bibinfo{year}{2017}), \bibinfo{pages}{2}.
\newblock


\bibitem[\protect\citeauthoryear{Dourish}{Dourish}{2004}]%
        {dourish2004action}
\bibfield{author}{\bibinfo{person}{Paul Dourish}.}
  \bibinfo{year}{2004}\natexlab{}.
\newblock \bibinfo{booktitle}{\emph{Where the action is: the foundations of
  embodied interaction}}.
\newblock \bibinfo{publisher}{MIT press}.
\newblock


\bibitem[\protect\citeauthoryear{Dourish, Finlay, Sengers, and Wright}{Dourish
  et~al\mbox{.}}{2004}]%
        {dourish2004reflective}
\bibfield{author}{\bibinfo{person}{Paul Dourish}, \bibinfo{person}{Janet
  Finlay}, \bibinfo{person}{Phoebe Sengers}, {and} \bibinfo{person}{Peter
  Wright}.} \bibinfo{year}{2004}\natexlab{}.
\newblock \showarticletitle{Reflective HCI: Towards a critical technical
  practice}. In \bibinfo{booktitle}{\emph{CHI'04 extended abstracts on Human
  factors in computing systems}}. \bibinfo{pages}{1727--1728}.
\newblock


\bibitem[\protect\citeauthoryear{Ehsan and Riedl}{Ehsan and Riedl}{2020}]%
        {ehsan2020human}
\bibfield{author}{\bibinfo{person}{Upol Ehsan} {and} \bibinfo{person}{Mark~O
  Riedl}.} \bibinfo{year}{2020}\natexlab{}.
\newblock \showarticletitle{Human-centered Explainable AI: Towards a Reflective
  Sociotechnical Approach}.
\newblock \bibinfo{journal}{\emph{arXiv preprint arXiv:2002.01092}}
  (\bibinfo{year}{2020}).
\newblock


\bibitem[\protect\citeauthoryear{Ehsan, Tambwekar, Chan, Harrison, and
  Riedl}{Ehsan et~al\mbox{.}}{2019}]%
        {ehsan2019automated}
\bibfield{author}{\bibinfo{person}{Upol Ehsan}, \bibinfo{person}{Pradyumna
  Tambwekar}, \bibinfo{person}{Larry Chan}, \bibinfo{person}{Brent Harrison},
  {and} \bibinfo{person}{Mark~O Riedl}.} \bibinfo{year}{2019}\natexlab{}.
\newblock \showarticletitle{Automated rationale generation: a technique for
  explainable AI and its effects on human perceptions}. In
  \bibinfo{booktitle}{\emph{Proceedings of the 24th International Conference on
  Intelligent User Interfaces}}. \bibinfo{pages}{263--274}.
\newblock


\bibitem[\protect\citeauthoryear{Erickson and Kellogg}{Erickson and
  Kellogg}{2003}]%
        {erickson2003social}
\bibfield{author}{\bibinfo{person}{Thomas Erickson} {and}
  \bibinfo{person}{Wendy~A Kellogg}.} \bibinfo{year}{2003}\natexlab{}.
\newblock \showarticletitle{Social translucence: using minimalist
  visualisations of social activity to support collective interaction}.
\newblock In \bibinfo{booktitle}{\emph{Designing information spaces: The social
  navigation approach}}. \bibinfo{publisher}{Springer},
  \bibinfo{pages}{17--41}.
\newblock


\bibitem[\protect\citeauthoryear{Ghai, Liao, Zhang, Bellamy, and Mueller}{Ghai
  et~al\mbox{.}}{2021}]%
        {Ghai2020XAL}
\bibfield{author}{\bibinfo{person}{Bhavya Ghai}, \bibinfo{person}{Q~Vera Liao},
  \bibinfo{person}{Yunfeng Zhang}, \bibinfo{person}{Rachel Bellamy}, {and}
  \bibinfo{person}{Klaus Mueller}.} \bibinfo{year}{2021}\natexlab{}.
\newblock \showarticletitle{Explainable Active Learning (XAL): Toward AI
  Explanations as Interfaces for Machine Teachers}.
\newblock \bibinfo{journal}{\emph{Proceedings of the ACM on Human-Computer
  Interaction}} \bibinfo{number}{CSCW} (\bibinfo{year}{2021}).
\newblock


\bibitem[\protect\citeauthoryear{Gilbert}{Gilbert}{2012}]%
        {gilbert2012designing}
\bibfield{author}{\bibinfo{person}{Eric Gilbert}.}
  \bibinfo{year}{2012}\natexlab{}.
\newblock \showarticletitle{Designing social translucence over social
  networks}. In \bibinfo{booktitle}{\emph{Proceedings of the SIGCHI Conference
  on Human Factors in Computing Systems}}. \bibinfo{pages}{2731--2740}.
\newblock


\bibitem[\protect\citeauthoryear{Gilpin, Bau, Yuan, Bajwa, Specter, and
  Kagal}{Gilpin et~al\mbox{.}}{2018}]%
        {gilpin2018explaining}
\bibfield{author}{\bibinfo{person}{Leilani~H Gilpin}, \bibinfo{person}{David
  Bau}, \bibinfo{person}{Ben~Z Yuan}, \bibinfo{person}{Ayesha Bajwa},
  \bibinfo{person}{Michael Specter}, {and} \bibinfo{person}{Lalana Kagal}.}
  \bibinfo{year}{2018}\natexlab{}.
\newblock \showarticletitle{Explaining explanations: An overview of
  interpretability of machine learning}. In \bibinfo{booktitle}{\emph{2018 IEEE
  5th International Conference on data science and advanced analytics (DSAA)}}.
  IEEE, \bibinfo{pages}{80--89}.
\newblock


\bibitem[\protect\citeauthoryear{Green and Viljoen}{Green and Viljoen}{2020}]%
        {green2020algorithmic}
\bibfield{author}{\bibinfo{person}{Ben Green} {and} \bibinfo{person}{Salom{\'e}
  Viljoen}.} \bibinfo{year}{2020}\natexlab{}.
\newblock \showarticletitle{Algorithmic realism: expanding the boundaries of
  algorithmic thought}. In \bibinfo{booktitle}{\emph{Proceedings of the 2020
  Conference on Fairness, Accountability, and Transparency}}.
  \bibinfo{pages}{19--31}.
\newblock


\bibitem[\protect\citeauthoryear{Grudin}{Grudin}{1988}]%
        {grudin1988cscw}
\bibfield{author}{\bibinfo{person}{Jonathan Grudin}.}
  \bibinfo{year}{1988}\natexlab{}.
\newblock \showarticletitle{Why CSCW applications fail: problems in the design
  and evaluationof organizational interfaces}. In
  \bibinfo{booktitle}{\emph{Proceedings of the 1988 ACM conference on
  Computer-supported cooperative work}}. \bibinfo{pages}{85--93}.
\newblock


\bibitem[\protect\citeauthoryear{Gunning}{Gunning}{2017}]%
        {gunning2017explainable}
\bibfield{author}{\bibinfo{person}{David Gunning}.}
  \bibinfo{year}{2017}\natexlab{}.
\newblock \showarticletitle{Explainable artificial intelligence (xai)}.
\newblock \bibinfo{journal}{\emph{Defense Advanced Research Projects Agency
  (DARPA), nd Web}}  \bibinfo{volume}{2} (\bibinfo{year}{2017}),
  \bibinfo{pages}{2}.
\newblock


\bibitem[\protect\citeauthoryear{Gutwin and Greenberg}{Gutwin and
  Greenberg}{2002}]%
        {gutwin2002descriptive}
\bibfield{author}{\bibinfo{person}{Carl Gutwin} {and} \bibinfo{person}{Saul
  Greenberg}.} \bibinfo{year}{2002}\natexlab{}.
\newblock \showarticletitle{A descriptive framework of workspace awareness for
  real-time groupware}.
\newblock \bibinfo{journal}{\emph{Computer supported cooperative work}}
  \bibinfo{volume}{11}, \bibinfo{number}{3-4} (\bibinfo{year}{2002}),
  \bibinfo{pages}{411--446}.
\newblock


\bibitem[\protect\citeauthoryear{Gutwin, Penner, and Schneider}{Gutwin
  et~al\mbox{.}}{2004}]%
        {gutwin2004group}
\bibfield{author}{\bibinfo{person}{Carl Gutwin}, \bibinfo{person}{Reagan
  Penner}, {and} \bibinfo{person}{Kevin Schneider}.}
  \bibinfo{year}{2004}\natexlab{}.
\newblock \showarticletitle{Group awareness in distributed software
  development}. In \bibinfo{booktitle}{\emph{Proceedings of the 2004 ACM
  conference on Computer supported cooperative work}}. \bibinfo{pages}{72--81}.
\newblock


\bibitem[\protect\citeauthoryear{Hao}{Hao}{2019}]%
        {hao2019jail}
\bibfield{author}{\bibinfo{person}{Karen Hao}.}
  \bibinfo{year}{2019}\natexlab{}.
\newblock \showarticletitle{AI is sending people to jail -- and getting it
  wrong}.
\newblock \bibinfo{journal}{\emph{MIT Technology Review}} (\bibinfo{date}{21
  January} \bibinfo{year}{2019}).
\newblock
\urldef\tempurl%
\url{https://www.technologyreview.com/s/612775/algorithms-
  criminal-justice-ai/}
\showURL{%
Retrieved 26-August-2019 from \tempurl}


\bibitem[\protect\citeauthoryear{Heider}{Heider}{1958}]%
        {heider1958psychology}
\bibfield{author}{\bibinfo{person}{Fritz Heider}.}
  \bibinfo{year}{1958}\natexlab{}.
\newblock \showarticletitle{The psychology of interpersonal relations Wiley}.
\newblock \bibinfo{journal}{\emph{New York}} (\bibinfo{year}{1958}).
\newblock


\bibitem[\protect\citeauthoryear{Hilton}{Hilton}{1996}]%
        {hilton1996mental}
\bibfield{author}{\bibinfo{person}{Denis~J Hilton}.}
  \bibinfo{year}{1996}\natexlab{}.
\newblock \showarticletitle{Mental models and causal explanation: Judgements of
  probable cause and explanatory relevance}.
\newblock \bibinfo{journal}{\emph{Thinking \& Reasoning}} \bibinfo{volume}{2},
  \bibinfo{number}{4} (\bibinfo{year}{1996}), \bibinfo{pages}{273--308}.
\newblock


\bibitem[\protect\citeauthoryear{Hind, Wei, Campbell, Codella, Dhurandhar,
  Mojsilovi{\'c}, Natesan~Ramamurthy, and Varshney}{Hind et~al\mbox{.}}{2019}]%
        {hind2019ted}
\bibfield{author}{\bibinfo{person}{Michael Hind}, \bibinfo{person}{Dennis Wei},
  \bibinfo{person}{Murray Campbell}, \bibinfo{person}{Noel~CF Codella},
  \bibinfo{person}{Amit Dhurandhar}, \bibinfo{person}{Aleksandra
  Mojsilovi{\'c}}, \bibinfo{person}{Karthikeyan Natesan~Ramamurthy}, {and}
  \bibinfo{person}{Kush~R Varshney}.} \bibinfo{year}{2019}\natexlab{}.
\newblock \showarticletitle{TED: Teaching AI to explain its decisions}. In
  \bibinfo{booktitle}{\emph{Proceedings of the 2019 AAAI/ACM Conference on AI,
  Ethics, and Society}}. \bibinfo{pages}{123--129}.
\newblock


\bibitem[\protect\citeauthoryear{Hoffman and Klein}{Hoffman and Klein}{2017}]%
        {hoffman2017explaining}
\bibfield{author}{\bibinfo{person}{Robert~R Hoffman} {and}
  \bibinfo{person}{Gary Klein}.} \bibinfo{year}{2017}\natexlab{}.
\newblock \showarticletitle{Explaining explanation, part 1: theoretical
  foundations}.
\newblock \bibinfo{journal}{\emph{IEEE Intelligent Systems}}
  \bibinfo{volume}{32}, \bibinfo{number}{3} (\bibinfo{year}{2017}),
  \bibinfo{pages}{68--73}.
\newblock


\bibitem[\protect\citeauthoryear{Hollingshead and Brandon}{Hollingshead and
  Brandon}{2003}]%
        {hollingshead2003potential}
\bibfield{author}{\bibinfo{person}{Andrea~B Hollingshead} {and}
  \bibinfo{person}{David~P Brandon}.} \bibinfo{year}{2003}\natexlab{}.
\newblock \showarticletitle{Potential benefits of communication in transactive
  memory systems}.
\newblock \bibinfo{journal}{\emph{Human communication research}}
  \bibinfo{volume}{29}, \bibinfo{number}{4} (\bibinfo{year}{2003}),
  \bibinfo{pages}{607--615}.
\newblock


\bibitem[\protect\citeauthoryear{Hong, Hullman, and Bertini}{Hong
  et~al\mbox{.}}{2020}]%
        {hong2020human}
\bibfield{author}{\bibinfo{person}{Sungsoo~Ray Hong}, \bibinfo{person}{Jessica
  Hullman}, {and} \bibinfo{person}{Enrico Bertini}.}
  \bibinfo{year}{2020}\natexlab{}.
\newblock \showarticletitle{Human Factors in Model Interpretability: Industry
  Practices, Challenges, and Needs}.
\newblock \bibinfo{journal}{\emph{Proceedings of the ACM on Human-Computer
  Interaction}} \bibinfo{volume}{4}, \bibinfo{number}{CSCW1}
  (\bibinfo{year}{2020}), \bibinfo{pages}{1--26}.
\newblock


\bibitem[\protect\citeauthoryear{Huang and Fu}{Huang and Fu}{2013}]%
        {huang2013don}
\bibfield{author}{\bibinfo{person}{Shih-Wen Huang} {and}
  \bibinfo{person}{Wai-Tat Fu}.} \bibinfo{year}{2013}\natexlab{}.
\newblock \showarticletitle{Don't hide in the crowd! Increasing social
  transparency between peer workers improves crowdsourcing outcomes}. In
  \bibinfo{booktitle}{\emph{Proceedings of the SIGCHI Conference on Human
  Factors in Computing Systems}}. \bibinfo{pages}{621--630}.
\newblock


\bibitem[\protect\citeauthoryear{Hume}{Hume}{2000}]%
        {hume2000enquiry}
\bibfield{author}{\bibinfo{person}{David Hume}.}
  \bibinfo{year}{2000}\natexlab{}.
\newblock \bibinfo{booktitle}{\emph{An enquiry concerning human understanding:
  A critical edition}}. Vol.~\bibinfo{volume}{3}.
\newblock \bibinfo{publisher}{Oxford University Press}.
\newblock


\bibitem[\protect\citeauthoryear{Hutchins}{Hutchins}{1991}]%
        {hutchins1991social}
\bibfield{author}{\bibinfo{person}{Edwin Hutchins}.}
  \bibinfo{year}{1991}\natexlab{}.
\newblock \showarticletitle{The social organization of distributed cognition.}
\newblock  (\bibinfo{year}{1991}).
\newblock


\bibitem[\protect\citeauthoryear{Jones, Artikis, and Pitt}{Jones
  et~al\mbox{.}}{2013}]%
        {jones2013design}
\bibfield{author}{\bibinfo{person}{Andrew~JI Jones}, \bibinfo{person}{Alexander
  Artikis}, {and} \bibinfo{person}{Jeremy Pitt}.}
  \bibinfo{year}{2013}\natexlab{}.
\newblock \showarticletitle{The design of intelligent socio-technical systems}.
\newblock \bibinfo{journal}{\emph{Artificial Intelligence Review}}
  \bibinfo{volume}{39}, \bibinfo{number}{1} (\bibinfo{year}{2013}),
  \bibinfo{pages}{5--20}.
\newblock


\bibitem[\protect\citeauthoryear{Kahneman, Slovic, Slovic, and
  Tversky}{Kahneman et~al\mbox{.}}{1982}]%
        {kahneman1982judgment}
\bibfield{author}{\bibinfo{person}{Daniel Kahneman},
  \bibinfo{person}{Stewart~Paul Slovic}, \bibinfo{person}{Paul Slovic}, {and}
  \bibinfo{person}{Amos Tversky}.} \bibinfo{year}{1982}\natexlab{}.
\newblock \bibinfo{booktitle}{\emph{Judgment under uncertainty: Heuristics and
  biases}}.
\newblock \bibinfo{publisher}{Cambridge university press}.
\newblock


\bibitem[\protect\citeauthoryear{Katell, Young, Dailey, Herman, Guetler, Tam,
  Bintz, Raz, and Krafft}{Katell et~al\mbox{.}}{2020}]%
        {katell2020toward}
\bibfield{author}{\bibinfo{person}{Michael Katell}, \bibinfo{person}{Meg
  Young}, \bibinfo{person}{Dharma Dailey}, \bibinfo{person}{Bernease Herman},
  \bibinfo{person}{Vivian Guetler}, \bibinfo{person}{Aaron Tam},
  \bibinfo{person}{Corinne Bintz}, \bibinfo{person}{Daniella Raz}, {and}
  \bibinfo{person}{PM Krafft}.} \bibinfo{year}{2020}\natexlab{}.
\newblock \showarticletitle{Toward situated interventions for algorithmic
  equity: lessons from the field}. In \bibinfo{booktitle}{\emph{Proceedings of
  the 2020 Conference on Fairness, Accountability, and Transparency}}.
  \bibinfo{pages}{45--55}.
\newblock


\bibitem[\protect\citeauthoryear{Kaur, Nori, Jenkins, Caruana, Wallach, and
  Wortman~Vaughan}{Kaur et~al\mbox{.}}{2020}]%
        {kaur2020interpreting}
\bibfield{author}{\bibinfo{person}{Harmanpreet Kaur}, \bibinfo{person}{Harsha
  Nori}, \bibinfo{person}{Samuel Jenkins}, \bibinfo{person}{Rich Caruana},
  \bibinfo{person}{Hanna Wallach}, {and} \bibinfo{person}{Jennifer
  Wortman~Vaughan}.} \bibinfo{year}{2020}\natexlab{}.
\newblock \showarticletitle{Interpreting Interpretability: Understanding Data
  Scientists' Use of Interpretability Tools for Machine Learning}. In
  \bibinfo{booktitle}{\emph{Proceedings of the 2020 CHI Conference on Human
  Factors in Computing Systems}}. \bibinfo{pages}{1--14}.
\newblock


\bibitem[\protect\citeauthoryear{Kim, Kong, Hong, and Karahalios}{Kim
  et~al\mbox{.}}{2020}]%
        {kim2020enriched}
\bibfield{author}{\bibinfo{person}{Jennifer~G Kim}, \bibinfo{person}{Ha-Kyung
  Kong}, \bibinfo{person}{Hwajung Hong}, {and} \bibinfo{person}{Karrie
  Karahalios}.} \bibinfo{year}{2020}\natexlab{}.
\newblock \showarticletitle{Enriched Social Translucence in Medical
  Crowdfunding}. In \bibinfo{booktitle}{\emph{Proceedings of the 2020 ACM
  Designing Interactive Systems Conference}}. \bibinfo{pages}{1465--1477}.
\newblock


\bibitem[\protect\citeauthoryear{Kramer}{Kramer}{1999}]%
        {kramer1999trust}
\bibfield{author}{\bibinfo{person}{Roderick~M Kramer}.}
  \bibinfo{year}{1999}\natexlab{}.
\newblock \showarticletitle{Trust and distrust in organizations: Emerging
  perspectives, enduring questions}.
\newblock \bibinfo{journal}{\emph{Annual review of psychology}}
  \bibinfo{volume}{50}, \bibinfo{number}{1} (\bibinfo{year}{1999}),
  \bibinfo{pages}{569--598}.
\newblock


\bibitem[\protect\citeauthoryear{Lai, Liu, and Tan}{Lai et~al\mbox{.}}{2020}]%
        {lai2020chicago}
\bibfield{author}{\bibinfo{person}{Vivian Lai}, \bibinfo{person}{Han Liu},
  {and} \bibinfo{person}{Chenhao Tan}.} \bibinfo{year}{2020}\natexlab{}.
\newblock \showarticletitle{" Why is' Chicago'deceptive?" Towards Building
  Model-Driven Tutorials for Humans}. In \bibinfo{booktitle}{\emph{Proceedings
  of the 2020 CHI Conference on Human Factors in Computing Systems}}.
  \bibinfo{pages}{1--13}.
\newblock


\bibitem[\protect\citeauthoryear{Lee, Kusbit, Kahng, Kim, Yuan, Chan, See,
  Noothigattu, Lee, Psomas, et~al\mbox{.}}{Lee et~al\mbox{.}}{2019}]%
        {lee2019webuildai}
\bibfield{author}{\bibinfo{person}{Min~Kyung Lee}, \bibinfo{person}{Daniel
  Kusbit}, \bibinfo{person}{Anson Kahng}, \bibinfo{person}{Ji~Tae Kim},
  \bibinfo{person}{Xinran Yuan}, \bibinfo{person}{Allissa Chan},
  \bibinfo{person}{Daniel See}, \bibinfo{person}{Ritesh Noothigattu},
  \bibinfo{person}{Siheon Lee}, \bibinfo{person}{Alexandros Psomas},
  {et~al\mbox{.}}} \bibinfo{year}{2019}\natexlab{}.
\newblock \showarticletitle{WeBuildAI: Participatory framework for algorithmic
  governance}.
\newblock \bibinfo{journal}{\emph{Proceedings of the ACM on Human-Computer
  Interaction}} \bibinfo{volume}{3}, \bibinfo{number}{CSCW}
  (\bibinfo{year}{2019}), \bibinfo{pages}{1--35}.
\newblock


\bibitem[\protect\citeauthoryear{Leonardi}{Leonardi}{2014}]%
        {leonardi2014social}
\bibfield{author}{\bibinfo{person}{Paul~M Leonardi}.}
  \bibinfo{year}{2014}\natexlab{}.
\newblock \showarticletitle{Social media, knowledge sharing, and innovation:
  Toward a theory of communication visibility}.
\newblock \bibinfo{journal}{\emph{Information systems research}}
  \bibinfo{volume}{25}, \bibinfo{number}{4} (\bibinfo{year}{2014}),
  \bibinfo{pages}{796--816}.
\newblock


\bibitem[\protect\citeauthoryear{Leonardi}{Leonardi}{2015}]%
        {leonardi2015ambient}
\bibfield{author}{\bibinfo{person}{Paul~M Leonardi}.}
  \bibinfo{year}{2015}\natexlab{}.
\newblock \showarticletitle{Ambient awareness and knowledge acquisition: using
  social media to learn ‘who knows what’and ‘who knows whom’}.
\newblock \bibinfo{journal}{\emph{Mis Quarterly}} \bibinfo{volume}{39},
  \bibinfo{number}{4} (\bibinfo{year}{2015}), \bibinfo{pages}{747--762}.
\newblock


\bibitem[\protect\citeauthoryear{Lewis}{Lewis}{1986}]%
        {lewis1986causal}
\bibfield{author}{\bibinfo{person}{David~K Lewis}.}
  \bibinfo{year}{1986}\natexlab{}.
\newblock \showarticletitle{Causal explanation}.
\newblock  (\bibinfo{year}{1986}).
\newblock


\bibitem[\protect\citeauthoryear{Liao, Gruen, and Miller}{Liao
  et~al\mbox{.}}{2020}]%
        {liao2020questioning}
\bibfield{author}{\bibinfo{person}{Q~Vera Liao}, \bibinfo{person}{Daniel
  Gruen}, {and} \bibinfo{person}{Sarah Miller}.}
  \bibinfo{year}{2020}\natexlab{}.
\newblock \showarticletitle{Questioning the AI: Informing Design Practices for
  Explainable AI User Experiences}. In \bibinfo{booktitle}{\emph{Proceedings of
  the SIGCHI Conference on Human Factors in Computing Systems}}. ACM.
\newblock


\bibitem[\protect\citeauthoryear{Lim and Dey}{Lim and Dey}{2010}]%
        {lim2010toolkit}
\bibfield{author}{\bibinfo{person}{Brian~Y Lim} {and} \bibinfo{person}{Anind~K
  Dey}.} \bibinfo{year}{2010}\natexlab{}.
\newblock \showarticletitle{Toolkit to support intelligibility in context-aware
  applications}. In \bibinfo{booktitle}{\emph{Proceedings of the 12th ACM
  international conference on Ubiquitous computing}}. \bibinfo{pages}{13--22}.
\newblock


\bibitem[\protect\citeauthoryear{Lim, Dey, and Avrahami}{Lim
  et~al\mbox{.}}{2009}]%
        {lim2009and}
\bibfield{author}{\bibinfo{person}{Brian~Y Lim}, \bibinfo{person}{Anind~K Dey},
  {and} \bibinfo{person}{Daniel Avrahami}.} \bibinfo{year}{2009}\natexlab{}.
\newblock \showarticletitle{Why and why not explanations improve the
  intelligibility of context-aware intelligent systems}. In
  \bibinfo{booktitle}{\emph{Proceedings of the SIGCHI Conference on Human
  Factors in Computing Systems}}. \bibinfo{pages}{2119--2128}.
\newblock


\bibitem[\protect\citeauthoryear{Lim, Yang, Abdul, and Wang}{Lim
  et~al\mbox{.}}{2019}]%
        {lim2019these}
\bibfield{author}{\bibinfo{person}{Brian~Y Lim}, \bibinfo{person}{Qian Yang},
  \bibinfo{person}{Ashraf~M Abdul}, {and} \bibinfo{person}{Danding Wang}.}
  \bibinfo{year}{2019}\natexlab{}.
\newblock \showarticletitle{Why these Explanations? Selecting Intelligibility
  Types for Explanation Goals.}. In \bibinfo{booktitle}{\emph{IUI Workshops}}.
\newblock


\bibitem[\protect\citeauthoryear{Lipton}{Lipton}{2018}]%
        {lipton2018mythos}
\bibfield{author}{\bibinfo{person}{Zachary~C Lipton}.}
  \bibinfo{year}{2018}\natexlab{}.
\newblock \showarticletitle{The mythos of model interpretability}.
\newblock \bibinfo{journal}{\emph{Queue}} \bibinfo{volume}{16},
  \bibinfo{number}{3} (\bibinfo{year}{2018}), \bibinfo{pages}{31--57}.
\newblock


\bibitem[\protect\citeauthoryear{Loftus, Tighe, Filiberto, Efron, Brakenridge,
  Mohr, Rashidi, Upchurch, and Bihorac}{Loftus et~al\mbox{.}}{2020}]%
        {loftus2020artificial}
\bibfield{author}{\bibinfo{person}{Tyler~J. Loftus},
  \bibinfo{person}{Patrick~J. Tighe}, \bibinfo{person}{Amanda~C. Filiberto},
  \bibinfo{person}{Philip~A. Efron}, \bibinfo{person}{Scott~C. Brakenridge},
  \bibinfo{person}{Alicia~M. Mohr}, \bibinfo{person}{Parisa Rashidi},
  \bibinfo{person}{Jr Upchurch, Gilbert~R.}, {and} \bibinfo{person}{Azra
  Bihorac}.} \bibinfo{year}{2020}\natexlab{}.
\newblock \showarticletitle{{Artificial Intelligence and Surgical
  Decision-making}}.
\newblock \bibinfo{journal}{\emph{JAMA Surgery}} \bibinfo{volume}{155},
  \bibinfo{number}{2} (\bibinfo{date}{02} \bibinfo{year}{2020}),
  \bibinfo{pages}{148--158}.
\newblock
\showISSN{2168-6254}
\urldef\tempurl%
\url{https://doi.org/10.1001/jamasurg.2019.4917}
\showDOI{\tempurl}
\showeprint{https://jamanetwork.com/journals/jamasurgery/articlepdf/2756311/jamasurgery\_loftus\_2019\_rv\_190007.pdf}


\bibitem[\protect\citeauthoryear{Lombrozo}{Lombrozo}{2011}]%
        {lombrozo2011instrumental}
\bibfield{author}{\bibinfo{person}{Tania Lombrozo}.}
  \bibinfo{year}{2011}\natexlab{}.
\newblock \showarticletitle{The instrumental value of explanations}.
\newblock \bibinfo{journal}{\emph{Philosophy Compass}} \bibinfo{volume}{6},
  \bibinfo{number}{8} (\bibinfo{year}{2011}), \bibinfo{pages}{539--551}.
\newblock


\bibitem[\protect\citeauthoryear{Lombrozo}{Lombrozo}{2012}]%
        {lombrozo2012explanation}
\bibfield{author}{\bibinfo{person}{Tania Lombrozo}.}
  \bibinfo{year}{2012}\natexlab{}.
\newblock \showarticletitle{Explanation and abductive inference.}
\newblock  (\bibinfo{year}{2012}).
\newblock


\bibitem[\protect\citeauthoryear{Makarius, Mukherjee, Fox, and Fox}{Makarius
  et~al\mbox{.}}{2020}]%
        {makarius2020rising}
\bibfield{author}{\bibinfo{person}{Erin~E Makarius}, \bibinfo{person}{Debmalya
  Mukherjee}, \bibinfo{person}{Joseph~D Fox}, {and} \bibinfo{person}{Alexa~K
  Fox}.} \bibinfo{year}{2020}\natexlab{}.
\newblock \showarticletitle{Rising with the machines: A sociotechnical
  framework for bringing artificial intelligence into the organization}.
\newblock \bibinfo{journal}{\emph{Journal of Business Research}}
  \bibinfo{volume}{120} (\bibinfo{year}{2020}), \bibinfo{pages}{262--273}.
\newblock


\bibitem[\protect\citeauthoryear{McDonald, Gokhman, and Zachry}{McDonald
  et~al\mbox{.}}{2012}]%
        {mcdonald2012building}
\bibfield{author}{\bibinfo{person}{David~W McDonald},
  \bibinfo{person}{Stephanie Gokhman}, {and} \bibinfo{person}{Mark Zachry}.}
  \bibinfo{year}{2012}\natexlab{}.
\newblock \showarticletitle{Building for social translucence: a domain analysis
  and prototype system}. In \bibinfo{booktitle}{\emph{Proceedings of the ACM
  2012 conference on computer supported cooperative work}}.
  \bibinfo{pages}{637--646}.
\newblock


\bibitem[\protect\citeauthoryear{Metzger and Flanagin}{Metzger and
  Flanagin}{2013}]%
        {metzger2013credibility}
\bibfield{author}{\bibinfo{person}{Miriam~J Metzger} {and}
  \bibinfo{person}{Andrew~J Flanagin}.} \bibinfo{year}{2013}\natexlab{}.
\newblock \showarticletitle{Credibility and trust of information in online
  environments: The use of cognitive heuristics}.
\newblock \bibinfo{journal}{\emph{Journal of pragmatics}}  \bibinfo{volume}{59}
  (\bibinfo{year}{2013}), \bibinfo{pages}{210--220}.
\newblock


\bibitem[\protect\citeauthoryear{Metzger, Flanagin, and Medders}{Metzger
  et~al\mbox{.}}{2010}]%
        {metzger2010social}
\bibfield{author}{\bibinfo{person}{Miriam~J Metzger}, \bibinfo{person}{Andrew~J
  Flanagin}, {and} \bibinfo{person}{Ryan~B Medders}.}
  \bibinfo{year}{2010}\natexlab{}.
\newblock \showarticletitle{Social and heuristic approaches to credibility
  evaluation online}.
\newblock \bibinfo{journal}{\emph{Journal of communication}}
  \bibinfo{volume}{60}, \bibinfo{number}{3} (\bibinfo{year}{2010}),
  \bibinfo{pages}{413--439}.
\newblock


\bibitem[\protect\citeauthoryear{Miller}{Miller}{2019}]%
        {miller2019explanation}
\bibfield{author}{\bibinfo{person}{Tim Miller}.}
  \bibinfo{year}{2019}\natexlab{}.
\newblock \showarticletitle{Explanation in artificial intelligence: Insights
  from the social sciences}.
\newblock \bibinfo{journal}{\emph{Artificial Intelligence}}
  \bibinfo{volume}{267} (\bibinfo{year}{2019}), \bibinfo{pages}{1--38}.
\newblock


\bibitem[\protect\citeauthoryear{Mittelstadt, Russell, and Wachter}{Mittelstadt
  et~al\mbox{.}}{2019}]%
        {mittelstadt2019explaining}
\bibfield{author}{\bibinfo{person}{Brent Mittelstadt}, \bibinfo{person}{Chris
  Russell}, {and} \bibinfo{person}{Sandra Wachter}.}
  \bibinfo{year}{2019}\natexlab{}.
\newblock \showarticletitle{Explaining explanations in AI}. In
  \bibinfo{booktitle}{\emph{Proceedings of the conference on fairness,
  accountability, and transparency}}. \bibinfo{pages}{279--288}.
\newblock


\bibitem[\protect\citeauthoryear{Mohamed, Png, and Isaac}{Mohamed
  et~al\mbox{.}}{2020}]%
        {mohamed2020decolonial}
\bibfield{author}{\bibinfo{person}{Shakir Mohamed},
  \bibinfo{person}{Marie-Therese Png}, {and} \bibinfo{person}{William Isaac}.}
  \bibinfo{year}{2020}\natexlab{}.
\newblock \showarticletitle{Decolonial AI: Decolonial Theory as Sociotechnical
  Foresight in Artificial Intelligence}.
\newblock \bibinfo{journal}{\emph{Philosophy \& Technology}}
  (\bibinfo{year}{2020}), \bibinfo{pages}{1--26}.
\newblock


\bibitem[\protect\citeauthoryear{Mohseni, Zarei, and Ragan}{Mohseni
  et~al\mbox{.}}{2018}]%
        {mohseni2018multidisciplinary}
\bibfield{author}{\bibinfo{person}{Sina Mohseni}, \bibinfo{person}{Niloofar
  Zarei}, {and} \bibinfo{person}{Eric~D Ragan}.}
  \bibinfo{year}{2018}\natexlab{}.
\newblock \showarticletitle{A Multidisciplinary Survey and Framework for Design
  and Evaluation of Explainable AI Systems}.
\newblock \bibinfo{journal}{\emph{arXiv}} (\bibinfo{year}{2018}),
  \bibinfo{pages}{arXiv--1811}.
\newblock


\bibitem[\protect\citeauthoryear{Moreland and Thompson}{Moreland and
  Thompson}{2006}]%
        {moreland2006transactive}
\bibfield{author}{\bibinfo{person}{Richard~L Moreland} {and} \bibinfo{person}{L
  Thompson}.} \bibinfo{year}{2006}\natexlab{}.
\newblock \showarticletitle{Transactive memory: Learning who knows what in work
  groups and organizations}.
\newblock \bibinfo{journal}{\emph{Small groups: Key readings}}
  \bibinfo{volume}{327} (\bibinfo{year}{2006}).
\newblock


\bibitem[\protect\citeauthoryear{Muller and Liao}{Muller and Liao}{[n.d.]}]%
        {muller2017exploring}
\bibfield{author}{\bibinfo{person}{Michael Muller} {and}
  \bibinfo{person}{Q~Vera Liao}.} \bibinfo{year}{[n.d.]}\natexlab{}.
\newblock \showarticletitle{Exploring AI Ethics and Values through
  Participatory Design Fictions}.
\newblock  (\bibinfo{year}{[n.\,d.]}).
\newblock


\bibitem[\protect\citeauthoryear{Murawski}{Murawski}{2019}]%
        {murawski2019mortgage}
\bibfield{author}{\bibinfo{person}{John Murawski}.}
  \bibinfo{year}{2019}\natexlab{}.
\newblock \showarticletitle{Mortgage Providers Look to AI to Process Home Loans
  Faster}.
\newblock \bibinfo{journal}{\emph{Wall Street Journal}} (\bibinfo{date}{18
  March} \bibinfo{year}{2019}).
\newblock
\urldef\tempurl%
\url{https://www.wsj.com/articles/mortgage-providers-look-to-ai-to-process-home-loans-faster-11552899212}
\showURL{%
Retrieved 16-September-2020 from \tempurl}


\bibitem[\protect\citeauthoryear{Nardi, Whittaker, and Schwarz}{Nardi
  et~al\mbox{.}}{2002}]%
        {nardi2002networkers}
\bibfield{author}{\bibinfo{person}{Bonnie~A Nardi}, \bibinfo{person}{Steve
  Whittaker}, {and} \bibinfo{person}{Heinrich Schwarz}.}
  \bibinfo{year}{2002}\natexlab{}.
\newblock \showarticletitle{NetWORKers and their activity in intensional
  networks}.
\newblock \bibinfo{journal}{\emph{Computer Supported Cooperative Work (CSCW)}}
  \bibinfo{volume}{11}, \bibinfo{number}{1-2} (\bibinfo{year}{2002}),
  \bibinfo{pages}{205--242}.
\newblock


\bibitem[\protect\citeauthoryear{Nguyen, Dabbish, and Kiesler}{Nguyen
  et~al\mbox{.}}{2015}]%
        {nguyen2015perverse}
\bibfield{author}{\bibinfo{person}{Duyen~T Nguyen}, \bibinfo{person}{Laura~A
  Dabbish}, {and} \bibinfo{person}{Sara Kiesler}.}
  \bibinfo{year}{2015}\natexlab{}.
\newblock \showarticletitle{The perverse effects of social transparency on
  online advice taking}. In \bibinfo{booktitle}{\emph{Proceedings of the 18th
  ACM Conference on Computer Supported Cooperative Work \& Social Computing}}.
  \bibinfo{pages}{207--217}.
\newblock


\bibitem[\protect\citeauthoryear{Pangburn}{Pangburn}{2019}]%
        {pangburn19schools}
\bibfield{author}{\bibinfo{person}{DJ Pangburn}.}
  \bibinfo{year}{2019}\natexlab{}.
\newblock \showarticletitle{Schools are using software to help pick who gets
  in. What could go wrong?}
\newblock \bibinfo{journal}{\emph{Fast Company}} (\bibinfo{date}{17 May}
  \bibinfo{year}{2019}).
\newblock
\urldef\tempurl%
\url{https://www.fastcompany.com/90342596/schools-are-quietly-turning-to-ai-to-help-pick-who-gets-in-what-could-go-wrong}
\showURL{%
Retrieved 16-September-2020 from \tempurl}


\bibitem[\protect\citeauthoryear{Poursabzi-Sangdeh, Goldstein, Hofman, Vaughan,
  and Wallach}{Poursabzi-Sangdeh et~al\mbox{.}}{2018}]%
        {poursabzi2018manipulating}
\bibfield{author}{\bibinfo{person}{Forough Poursabzi-Sangdeh},
  \bibinfo{person}{Daniel~G Goldstein}, \bibinfo{person}{Jake~M Hofman},
  \bibinfo{person}{Jennifer~Wortman Vaughan}, {and} \bibinfo{person}{Hanna
  Wallach}.} \bibinfo{year}{2018}\natexlab{}.
\newblock \showarticletitle{Manipulating and measuring model interpretability}.
\newblock \bibinfo{journal}{\emph{arXiv preprint arXiv:1802.07810}}
  (\bibinfo{year}{2018}).
\newblock


\bibitem[\protect\citeauthoryear{Rader, Cotter, and Cho}{Rader
  et~al\mbox{.}}{2018}]%
        {rader2018explanations}
\bibfield{author}{\bibinfo{person}{Emilee Rader}, \bibinfo{person}{Kelley
  Cotter}, {and} \bibinfo{person}{Janghee Cho}.}
  \bibinfo{year}{2018}\natexlab{}.
\newblock \showarticletitle{Explanations as mechanisms for supporting
  algorithmic transparency}. In \bibinfo{booktitle}{\emph{Proceedings of the
  2018 CHI conference on human factors in computing systems}}.
  \bibinfo{pages}{1--13}.
\newblock


\bibitem[\protect\citeauthoryear{Ras, van Gerven, and Haselager}{Ras
  et~al\mbox{.}}{2018}]%
        {ras2018explanation}
\bibfield{author}{\bibinfo{person}{Gabri{\"e}lle Ras}, \bibinfo{person}{Marcel
  van Gerven}, {and} \bibinfo{person}{Pim Haselager}.}
  \bibinfo{year}{2018}\natexlab{}.
\newblock \showarticletitle{Explanation methods in deep learning: Users,
  values, concerns and challenges}.
\newblock In \bibinfo{booktitle}{\emph{Explainable and Interpretable Models in
  Computer Vision and Machine Learning}}. \bibinfo{publisher}{Springer},
  \bibinfo{pages}{19--36}.
\newblock


\bibitem[\protect\citeauthoryear{Resnick, Kuwabara, Zeckhauser, and
  Friedman}{Resnick et~al\mbox{.}}{2000}]%
        {resnick2000reputation}
\bibfield{author}{\bibinfo{person}{Paul Resnick}, \bibinfo{person}{Ko
  Kuwabara}, \bibinfo{person}{Richard Zeckhauser}, {and} \bibinfo{person}{Eric
  Friedman}.} \bibinfo{year}{2000}\natexlab{}.
\newblock \showarticletitle{Reputation systems}.
\newblock \bibinfo{journal}{\emph{Commun. ACM}} \bibinfo{volume}{43},
  \bibinfo{number}{12} (\bibinfo{year}{2000}), \bibinfo{pages}{45--48}.
\newblock


\bibitem[\protect\citeauthoryear{Rosson and Carroll}{Rosson and
  Carroll}{2009}]%
        {rosson2009scenario}
\bibfield{author}{\bibinfo{person}{Mary~Beth Rosson} {and}
  \bibinfo{person}{John~M Carroll}.} \bibinfo{year}{2009}\natexlab{}.
\newblock \showarticletitle{Scenario based design}.
\newblock \bibinfo{journal}{\emph{Human-computer interaction. boca raton, FL}}
  (\bibinfo{year}{2009}), \bibinfo{pages}{145--162}.
\newblock


\bibitem[\protect\citeauthoryear{{\v{S}}abanovi{\'c}}{{\v{S}}abanovi{\'c}}{2010}]%
        {sabanovic2010robots}
\bibfield{author}{\bibinfo{person}{Selma {\v{S}}abanovi{\'c}}.}
  \bibinfo{year}{2010}\natexlab{}.
\newblock \showarticletitle{Robots in society, society in robots}.
\newblock \bibinfo{journal}{\emph{International Journal of Social Robotics}}
  \bibinfo{volume}{2}, \bibinfo{number}{4} (\bibinfo{year}{2010}),
  \bibinfo{pages}{439--450}.
\newblock


\bibitem[\protect\citeauthoryear{S{\'a}nchez-Monedero, Dencik, and
  Edwards}{S{\'a}nchez-Monedero et~al\mbox{.}}{2020}]%
        {sanchez2020does}
\bibfield{author}{\bibinfo{person}{Javier S{\'a}nchez-Monedero},
  \bibinfo{person}{Lina Dencik}, {and} \bibinfo{person}{Lilian Edwards}.}
  \bibinfo{year}{2020}\natexlab{}.
\newblock \showarticletitle{What does it mean to'solve'the problem of
  discrimination in hiring? social, technical and legal perspectives from the
  UK on automated hiring systems}. In \bibinfo{booktitle}{\emph{Proceedings of
  the 2020 Conference on Fairness, Accountability, and Transparency}}.
  \bibinfo{pages}{458--468}.
\newblock


\bibitem[\protect\citeauthoryear{Selbst, Boyd, Friedler, Venkatasubramanian,
  and Vertesi}{Selbst et~al\mbox{.}}{2019}]%
        {selbst2019fairness}
\bibfield{author}{\bibinfo{person}{Andrew~D Selbst}, \bibinfo{person}{Danah
  Boyd}, \bibinfo{person}{Sorelle~A Friedler}, \bibinfo{person}{Suresh
  Venkatasubramanian}, {and} \bibinfo{person}{Janet Vertesi}.}
  \bibinfo{year}{2019}\natexlab{}.
\newblock \showarticletitle{Fairness and abstraction in sociotechnical
  systems}. In \bibinfo{booktitle}{\emph{Proceedings of the Conference on
  Fairness, Accountability, and Transparency}}. \bibinfo{pages}{59--68}.
\newblock


\bibitem[\protect\citeauthoryear{Sengers, Boehner, David, and Kaye}{Sengers
  et~al\mbox{.}}{2005}]%
        {sengers2005reflective}
\bibfield{author}{\bibinfo{person}{Phoebe Sengers}, \bibinfo{person}{Kirsten
  Boehner}, \bibinfo{person}{Shay David}, {and} \bibinfo{person}{Joseph'Jofish'
  Kaye}.} \bibinfo{year}{2005}\natexlab{}.
\newblock \showarticletitle{Reflective design}. In
  \bibinfo{booktitle}{\emph{Proceedings of the 4th decennial conference on
  Critical computing: between sense and sensibility}}. \bibinfo{pages}{49--58}.
\newblock


\bibitem[\protect\citeauthoryear{Shneiderman}{Shneiderman}{2020}]%
        {shneiderman2020human}
\bibfield{author}{\bibinfo{person}{Ben Shneiderman}.}
  \bibinfo{year}{2020}\natexlab{}.
\newblock \showarticletitle{Human-centered artificial intelligence: Reliable,
  safe \& trustworthy}.
\newblock \bibinfo{journal}{\emph{International Journal of Human--Computer
  Interaction}} \bibinfo{volume}{36}, \bibinfo{number}{6}
  (\bibinfo{year}{2020}), \bibinfo{pages}{495--504}.
\newblock


\bibitem[\protect\citeauthoryear{Smith-Renner, Fan, Birchfield, Wu,
  Boyd-Graber, Weld, and Findlater}{Smith-Renner et~al\mbox{.}}{2020}]%
        {smith2020no}
\bibfield{author}{\bibinfo{person}{Alison Smith-Renner}, \bibinfo{person}{Ron
  Fan}, \bibinfo{person}{Melissa Birchfield}, \bibinfo{person}{Tongshuang Wu},
  \bibinfo{person}{Jordan Boyd-Graber}, \bibinfo{person}{Daniel~S Weld}, {and}
  \bibinfo{person}{Leah Findlater}.} \bibinfo{year}{2020}\natexlab{}.
\newblock \showarticletitle{No Explainability without Accountability: An
  Empirical Study of Explanations and Feedback in Interactive ML}. In
  \bibinfo{booktitle}{\emph{Proceedings of the 2020 CHI Conference on Human
  Factors in Computing Systems}}. \bibinfo{pages}{1--13}.
\newblock


\bibitem[\protect\citeauthoryear{Spinner, Schlegel, Sch{\"a}fer, and
  El-Assady}{Spinner et~al\mbox{.}}{2019}]%
        {spinner2019explainer}
\bibfield{author}{\bibinfo{person}{Thilo Spinner}, \bibinfo{person}{Udo
  Schlegel}, \bibinfo{person}{Hanna Sch{\"a}fer}, {and}
  \bibinfo{person}{Mennatallah El-Assady}.} \bibinfo{year}{2019}\natexlab{}.
\newblock \showarticletitle{explAIner: A visual analytics framework for
  interactive and explainable machine learning}.
\newblock \bibinfo{journal}{\emph{IEEE transactions on visualization and
  computer graphics}} \bibinfo{volume}{26}, \bibinfo{number}{1}
  (\bibinfo{year}{2019}), \bibinfo{pages}{1064--1074}.
\newblock


\bibitem[\protect\citeauthoryear{Star and Strauss}{Star and Strauss}{1999}]%
        {star1999layers}
\bibfield{author}{\bibinfo{person}{Susan~Leigh Star} {and}
  \bibinfo{person}{Anselm Strauss}.} \bibinfo{year}{1999}\natexlab{}.
\newblock \showarticletitle{Layers of silence, arenas of voice: The ecology of
  visible and invisible work}.
\newblock \bibinfo{journal}{\emph{Computer supported cooperative work (CSCW)}}
  \bibinfo{volume}{8}, \bibinfo{number}{1-2} (\bibinfo{year}{1999}),
  \bibinfo{pages}{9--30}.
\newblock


\bibitem[\protect\citeauthoryear{Strauss and Corbin}{Strauss and
  Corbin}{1994}]%
        {strauss1994grounded}
\bibfield{author}{\bibinfo{person}{Anselm Strauss} {and}
  \bibinfo{person}{Juliet Corbin}.} \bibinfo{year}{1994}\natexlab{}.
\newblock \showarticletitle{Grounded theory methodology}.
\newblock \bibinfo{journal}{\emph{Handbook of qualitative research}}
  \bibinfo{volume}{17}, \bibinfo{number}{1} (\bibinfo{year}{1994}),
  \bibinfo{pages}{273--285}.
\newblock


\bibitem[\protect\citeauthoryear{Stuart, Dabbish, Kiesler, Kinnaird, and
  Kang}{Stuart et~al\mbox{.}}{2012}]%
        {stuart2012social}
\bibfield{author}{\bibinfo{person}{H~Colleen Stuart}, \bibinfo{person}{Laura
  Dabbish}, \bibinfo{person}{Sara Kiesler}, \bibinfo{person}{Peter Kinnaird},
  {and} \bibinfo{person}{Ruogu Kang}.} \bibinfo{year}{2012}\natexlab{}.
\newblock \showarticletitle{Social transparency in networked information
  exchange: a theoretical framework}. In \bibinfo{booktitle}{\emph{Proceedings
  of the ACM 2012 conference on Computer Supported Cooperative Work}}.
  \bibinfo{pages}{451--460}.
\newblock


\bibitem[\protect\citeauthoryear{Stumpf, Bussone, and O’sullivan}{Stumpf
  et~al\mbox{.}}{2016}]%
        {stumpf2016explanations}
\bibfield{author}{\bibinfo{person}{Simone Stumpf}, \bibinfo{person}{Adrian
  Bussone}, {and} \bibinfo{person}{Dympna O’sullivan}.}
  \bibinfo{year}{2016}\natexlab{}.
\newblock \showarticletitle{Explanations considered harmful? user interactions
  with machine learning systems}. In \bibinfo{booktitle}{\emph{ACM SIGCHI
  Workshop on Human-Centered Machine Learning}}.
\newblock


\bibitem[\protect\citeauthoryear{Suchman}{Suchman}{1995}]%
        {suchman1995making}
\bibfield{author}{\bibinfo{person}{Lucy Suchman}.}
  \bibinfo{year}{1995}\natexlab{}.
\newblock \showarticletitle{Making work visible}.
\newblock \bibinfo{journal}{\emph{Commun. ACM}} \bibinfo{volume}{38},
  \bibinfo{number}{9} (\bibinfo{year}{1995}), \bibinfo{pages}{56--64}.
\newblock


\bibitem[\protect\citeauthoryear{Suchman}{Suchman}{1987}]%
        {suchman1987plans}
\bibfield{author}{\bibinfo{person}{Lucy~A Suchman}.}
  \bibinfo{year}{1987}\natexlab{}.
\newblock \bibinfo{booktitle}{\emph{Plans and situated actions: The problem of
  human-machine communication}}.
\newblock \bibinfo{publisher}{Cambridge university press}.
\newblock


\bibitem[\protect\citeauthoryear{Sundar}{Sundar}{2008}]%
        {sundar2008main}
\bibfield{author}{\bibinfo{person}{S~Shyam Sundar}.}
  \bibinfo{year}{2008}\natexlab{}.
\newblock \bibinfo{booktitle}{\emph{The MAIN model: A heuristic approach to
  understanding technology effects on credibility}}.
\newblock \bibinfo{publisher}{MacArthur Foundation Digital Media and Learning
  Initiative}.
\newblock


\bibitem[\protect\citeauthoryear{Suresh and Guttag}{Suresh and Guttag}{2019}]%
        {suresh2019framework}
\bibfield{author}{\bibinfo{person}{Harini Suresh} {and} \bibinfo{person}{John~V
  Guttag}.} \bibinfo{year}{2019}\natexlab{}.
\newblock \showarticletitle{A framework for understanding unintended
  consequences of machine learning}.
\newblock \bibinfo{journal}{\emph{arXiv preprint arXiv:1901.10002}}
  (\bibinfo{year}{2019}).
\newblock


\bibitem[\protect\citeauthoryear{Vaughan and Wallach}{Vaughan and
  Wallach}{[n.d.]}]%
        {vaughan20201}
\bibfield{author}{\bibinfo{person}{Jennifer~Wortman Vaughan} {and}
  \bibinfo{person}{Hanna Wallach}.} \bibinfo{year}{[n.d.]}\natexlab{}.
\newblock \showarticletitle{1 A Human-Centered Agenda for Intelligible Machine
  Learning}.
\newblock  (\bibinfo{year}{[n.\,d.]}).
\newblock


\bibitem[\protect\citeauthoryear{Wang, Yang, Abdul, and Lim}{Wang
  et~al\mbox{.}}{2019}]%
        {wang2019designing}
\bibfield{author}{\bibinfo{person}{Danding Wang}, \bibinfo{person}{Qian Yang},
  \bibinfo{person}{Ashraf Abdul}, {and} \bibinfo{person}{Brian~Y Lim}.}
  \bibinfo{year}{2019}\natexlab{}.
\newblock \showarticletitle{Designing theory-driven user-centric explainable
  AI}. In \bibinfo{booktitle}{\emph{Proceedings of the 2019 CHI conference on
  human factors in computing systems}}. \bibinfo{pages}{1--15}.
\newblock


\bibitem[\protect\citeauthoryear{Wegner, Erber, and Raymond}{Wegner
  et~al\mbox{.}}{1991}]%
        {wegner1991transactive}
\bibfield{author}{\bibinfo{person}{Daniel~M Wegner}, \bibinfo{person}{Ralph
  Erber}, {and} \bibinfo{person}{Paula Raymond}.}
  \bibinfo{year}{1991}\natexlab{}.
\newblock \showarticletitle{Transactive memory in close relationships.}
\newblock \bibinfo{journal}{\emph{Journal of personality and social
  psychology}} \bibinfo{volume}{61}, \bibinfo{number}{6}
  (\bibinfo{year}{1991}), \bibinfo{pages}{923}.
\newblock


\bibitem[\protect\citeauthoryear{Weick and Roberts}{Weick and Roberts}{1993}]%
        {weick1993collective}
\bibfield{author}{\bibinfo{person}{Karl~E Weick} {and}
  \bibinfo{person}{Karlene~H Roberts}.} \bibinfo{year}{1993}\natexlab{}.
\newblock \showarticletitle{Collective mind in organizations: Heedful
  interrelating on flight decks}.
\newblock \bibinfo{journal}{\emph{Administrative science quarterly}}
  (\bibinfo{year}{1993}), \bibinfo{pages}{357--381}.
\newblock


\bibitem[\protect\citeauthoryear{Weld and Bansal}{Weld and Bansal}{2019}]%
        {weld2019challenge}
\bibfield{author}{\bibinfo{person}{Daniel~S Weld} {and} \bibinfo{person}{Gagan
  Bansal}.} \bibinfo{year}{2019}\natexlab{}.
\newblock \showarticletitle{The challenge of crafting intelligible
  intelligence}.
\newblock \bibinfo{journal}{\emph{Commun. ACM}} \bibinfo{volume}{62},
  \bibinfo{number}{6} (\bibinfo{year}{2019}), \bibinfo{pages}{70--79}.
\newblock


\bibitem[\protect\citeauthoryear{Wilkenfeld and Lombrozo}{Wilkenfeld and
  Lombrozo}{2015}]%
        {wilkenfeld2015inference}
\bibfield{author}{\bibinfo{person}{Daniel~A Wilkenfeld} {and}
  \bibinfo{person}{Tania Lombrozo}.} \bibinfo{year}{2015}\natexlab{}.
\newblock \showarticletitle{Inference to the best explanation (IBE) versus
  explaining for the best inference (EBI)}.
\newblock \bibinfo{journal}{\emph{Science \& Education}} \bibinfo{volume}{24},
  \bibinfo{number}{9-10} (\bibinfo{year}{2015}), \bibinfo{pages}{1059--1077}.
\newblock


\bibitem[\protect\citeauthoryear{Wolf and Blomberg}{Wolf and Blomberg}{2019}]%
        {wolf2019evaluating}
\bibfield{author}{\bibinfo{person}{Christine Wolf} {and}
  \bibinfo{person}{Jeanette Blomberg}.} \bibinfo{year}{2019}\natexlab{}.
\newblock \showarticletitle{Evaluating the promise of human-algorithm
  collaborations in everyday work practices}.
\newblock \bibinfo{journal}{\emph{Proceedings of the ACM on Human-Computer
  Interaction}} \bibinfo{volume}{3}, \bibinfo{number}{CSCW}
  (\bibinfo{year}{2019}), \bibinfo{pages}{1--23}.
\newblock


\bibitem[\protect\citeauthoryear{Yang, Huang, Scholtz, and Arendt}{Yang
  et~al\mbox{.}}{2020}]%
        {yang2020visual}
\bibfield{author}{\bibinfo{person}{Fumeng Yang}, \bibinfo{person}{Zhuanyi
  Huang}, \bibinfo{person}{Jean Scholtz}, {and} \bibinfo{person}{Dustin~L
  Arendt}.} \bibinfo{year}{2020}\natexlab{}.
\newblock \showarticletitle{How do visual explanations foster end users'
  appropriate trust in machine learning?}. In
  \bibinfo{booktitle}{\emph{Proceedings of the 25th International Conference on
  Intelligent User Interfaces}}. \bibinfo{pages}{189--201}.
\newblock


\bibitem[\protect\citeauthoryear{Yang, Steinfeld, and Zimmerman}{Yang
  et~al\mbox{.}}{2019}]%
        {yang2019unremarkable}
\bibfield{author}{\bibinfo{person}{Qian Yang}, \bibinfo{person}{Aaron
  Steinfeld}, {and} \bibinfo{person}{John Zimmerman}.}
  \bibinfo{year}{2019}\natexlab{}.
\newblock \showarticletitle{Unremarkable ai: Fitting intelligent decision
  support into critical, clinical decision-making processes}. In
  \bibinfo{booktitle}{\emph{Proceedings of the 2019 CHI Conference on Human
  Factors in Computing Systems}}. \bibinfo{pages}{1--11}.
\newblock


\bibitem[\protect\citeauthoryear{Yang, Zimmerman, Steinfeld, Carey, and
  Antaki}{Yang et~al\mbox{.}}{2016}]%
        {yang2016investigating}
\bibfield{author}{\bibinfo{person}{Qian Yang}, \bibinfo{person}{John
  Zimmerman}, \bibinfo{person}{Aaron Steinfeld}, \bibinfo{person}{Lisa Carey},
  {and} \bibinfo{person}{James~F Antaki}.} \bibinfo{year}{2016}\natexlab{}.
\newblock \showarticletitle{Investigating the heart pump implant decision
  process: opportunities for decision support tools to help}. In
  \bibinfo{booktitle}{\emph{Proceedings of the 2016 CHI Conference on Human
  Factors in Computing Systems}}. \bibinfo{pages}{4477--4488}.
\newblock


\bibitem[\protect\citeauthoryear{Yoo and Kanawattanachai}{Yoo and
  Kanawattanachai}{2001}]%
        {yoo2001developments}
\bibfield{author}{\bibinfo{person}{Youngjin Yoo} {and} \bibinfo{person}{Prasert
  Kanawattanachai}.} \bibinfo{year}{2001}\natexlab{}.
\newblock \showarticletitle{Developments of transactive memory systems and
  collective mind in virtual teams}.
\newblock \bibinfo{journal}{\emph{International Journal of Organizational
  Analysis}} \bibinfo{volume}{9}, \bibinfo{number}{2} (\bibinfo{year}{2001}),
  \bibinfo{pages}{187--208}.
\newblock


\bibitem[\protect\citeauthoryear{Zhang, Liao, and Bellamy}{Zhang
  et~al\mbox{.}}{2020}]%
        {zhang2020effect}
\bibfield{author}{\bibinfo{person}{Yunfeng Zhang}, \bibinfo{person}{Q~Vera
  Liao}, {and} \bibinfo{person}{Rachel~KE Bellamy}.}
  \bibinfo{year}{2020}\natexlab{}.
\newblock \showarticletitle{Effect of Confidence and Explanation on Accuracy
  and Trust Calibration in AI-Assisted Decision Making}. In
  \bibinfo{booktitle}{\emph{Proceedings of the Conference on Fairness,
  Accountability, and Transparency}}. ACM.
\newblock


\bibitem[\protect\citeauthoryear{Zhu, Yu, Halfaker, and Terveen}{Zhu
  et~al\mbox{.}}{2018}]%
        {zhu2018value}
\bibfield{author}{\bibinfo{person}{Haiyi Zhu}, \bibinfo{person}{Bowen Yu},
  \bibinfo{person}{Aaron Halfaker}, {and} \bibinfo{person}{Loren Terveen}.}
  \bibinfo{year}{2018}\natexlab{}.
\newblock \showarticletitle{Value-sensitive algorithm design: Method, case
  study, and lessons}.
\newblock \bibinfo{journal}{\emph{Proceedings of the ACM on Human-Computer
  Interaction}} \bibinfo{volume}{2}, \bibinfo{number}{CSCW}
  (\bibinfo{year}{2018}), \bibinfo{pages}{1--23}.
\newblock


\end{thebibliography}

\end{document}